\RequirePackage{etex}
\documentclass[longauth]{aa}  
\usepackage[utf8]{inputenc}

\usepackage{lipsum}
\usepackage{subcaption}         
\usepackage{lscape}             
\usepackage{placeins}           
\usepackage{graphicx,xspace}
\usepackage{amssymb}
\usepackage{hyperref}
\usepackage{dirtytalk}
\usepackage[dvipsnames]{xcolor}
\hypersetup{colorlinks=true,linkcolor=blue,citecolor=blue,filecolor=blue,urlcolor=blue}

\usepackage[varg]{txfonts}
\usepackage{threeparttable}
\usepackage{adjustbox}
\usepackage{float}

\newcommand{\erosita}{{eROSITA}}
\newcommand{\gaia}{{\it Gaia}}
\newcommand{\rosat}{{\it ROSAT}}
\newcommand{\SRG}{{\it SRG}}

\newcommand{\tess}{{\it TESS}}
\newcommand{\lisa}{{\it LISA}}

\begin{document}

   \title{SRG/\erosita\ X-ray selected cataclysmic variable candidates observed in SDSS-V DR20}

   \subtitle{}

 \author{J. Brink
          \thanks{jbrink@aip.de}\inst{1,2}
          \and
          A.D. Schwope\inst{1}
          \and
          K.G. Pradeep\inst{1,2}
          \and
          T. Dwelly\inst{3}
          \and
          S.F. Anderson\inst{4}
          \and
          C. Andonie\inst{3}
          \and
          S. Aviram \inst{5}
          \and
          C. Aydar\inst{3}
          \and
          W. N. Brandt\inst{6,7,8}
          \and
          J. Buchner\inst{3}
          \and
          S. Demasi\inst{4}
          \and
          M. Eracleous\inst{6}
          \and
          S. Friedrich\inst{3}
          \and
          B.T. Gänsicke\inst{9}
          \and
          F. Haberl\inst{3}
          \and
          P.B. Hall\inst{10}
          \and
          J.J. Hermes\inst{11}
          \and
          S. Hernández-Díaz\inst{12}
          \and
          D. Homan\inst{1,13}
          \and
          K. Inight\inst{9}
          \and
          T. Kupfer\inst{14}
          \and
          J. Kurpas\inst{1}
          \and
          C. Maitra\inst{3,15}
          \and    
          A. Merloni\inst{3}
          \and
          D. Muñoz-Giraldo\inst{12}
          \and
          M.K.M. Prescott\inst{16}        
          \and
          A.L. Rankine\inst{17}
          \and
          A.C. Rodriguez\inst{18}
          \and
          W. Roster\inst{3}
          \and
          S. Saeedi\inst{19}
          \and
          M. Salvato\inst{3}
          \and
          D.P. Schneider\inst{6,7}
          \and
          M.R. Schreiber\inst{20}
          \and
          B. Stelzer\inst{12}
          \and
          K. Szekerczes \inst{6,7}
          \and
          G. Tovmasian\inst{21}
          \and
          B. Trakhtenbrot\inst{5}
          \and
          Q. Wu\inst{22}
        }

   \institute{Leibniz-Institut für Astrophysik Potsdam (AIP), An der Sternwarte 16, 14482 Potsdam, Germany
         \and 
             Institute for Physics and Astronomy, University of Potsdam, Karl-Liebknecht-Str. 24/25, 14476 Potsdam, Germany
        \and 
            Max-Planck-Institut für extraterrestrische Physik, Giessenbachstrasse 1, 85748 Garching, Germany
        \and 
            Department of Astronomy, University of Washington, Box 351580, Seattle, WA 98195.
        \and 
            School of Physics and Astronomy, Tel Aviv University, Tel Aviv 69978, Israel
        \and 
             Department of Astronomy and Astrophysics, 525 Davey Laboratory, The Pennsylvania State University, University Park, PA 16802, USA
        \and 
            Institute for Gravitation and the Cosmos, 104 Davey Laboratory, The Pennsylvania State University, University Park, PA 16802, USA
        \and 
            Department of Physics, 104 Davey Laboratory, The Pennsylvania State University, University Park, PA 16802, USA
        \and 
            Department of Physics, University of Warwick, Coventry CV4 7AL, UK
        \and 
            Department of Physics and Astronomy, York University, 4700 Keele Street, Toronto, ON M3J 1P3, Canada
        \and 
            Department of Astronomy and Institute for Astrophysical Research, Boston University, 725 Commonwealth Avenue, Boston, MA 02215, USA
        \and 
            Institut für Astronomie und Astrophysik, Eberhard Karls Universität Tübingen, Sand 1, 72076 Tübingen, Germany
        \and 
            Institute of Astronomy, University of Cambridge, Madingley Road, Cambridge, CB3 0HA, United Kingdom
        \and 
            Hamburger Sternwarte, University of Hamburg, Gojenbergsweg 112, 21029 Hamburg, Germany
        \and 
            Inter-University Centre for Astronomy and Astrophysics (IUCAA), Ganeshkhind, Pune 411007, India
        \and 
            Department of Astronomy, New Mexico State University, Las Cruces, NM 88003, USA
        \and 
            Institute for Astronomy, University of Edinburgh, Royal Observatory, Edinburgh EH9 3HJ, UK
        \and 
            Department of Astronomy, California Institute of Technology, 1200 East California Blvd, Pasadena, CA 91125, USA
        \and 
            Dr. Karl Remeis-Sternwarte and Erlangen Centre for Astroparticle Physics, Friedrich-Alexander Universität Erlangen-Nürnberg, Sternwartstraße 7, 96049 Bamberg, Germany
        \and 
            Departamento de Física, Universidad Técnica Federico Santa María, Av. España 1680, Valparaíso, Chile
        \and 
            Universidad Nacional Autónoma de México, Instituto de Astronomía, A.P. 70-264, 04510 Ciudad de México, México
        \and 
            Department of Astronomy, University of Illinois at Urbana-Champaign, Urbana, IL 61801, USA
        }

   \date{Received xx, 20XX}

 
\abstract
{We report on the spectroscopic observations obtained during SDSS-V DR20 of accreting compact binaries (ACB), specifically the cataclysmic variables (CVs), that were identified as likely compact binary candidates from the SRG (Spectrum Roentgen Gamma) \erosita\ eRASS1 and eRASS:3 observations.}
{Our primary aim is to obtain a complete inventory of all CVs that were detected in eRASS1 and eRASS:3, with the goal to help better understand close-binary evolution, the population density, and demographics of these systems in the Galaxy. Previous population studies had their respective limitations, with volume-limited samples suffering from low-number statistics, while magnitude-limited observations were biased to only the brighter systems. All CVs are X-ray emitters, \erosita, therefore, presents a unique opportunity to identify CVs based on their X-ray emission, with eRASS1 reaching F$_X$ > $ 5 \times 10^{-14}$ erg cm$^{-2}$ s$^{-1}$ and eRASS:3 F$_X$ $\sim 2.7 \times 10^{-14}$ erg cm$^{-2}$ s$^{-1}$ in the 0.2--2.3 keV band. In fact, given these sensitivities, we expect to find most X-ray active CVs within 500 pc, and magnetic CVs to several kpc, in the \erosita\ data sample.} 
{Using X-ray data from the first \erosita\ all-sky survey, eRASS1, as well as the combined stack of the first 3 surveys, eRASS:3, together with optical data from \gaia\/, three different approaches were explored to identify the Gaia optical counterpart to the eROSITA X-ray source of the likely ACB. From this 89,337 unique candidates were identified and were submitted in three different cartons, i.e. groups of targets identified by specific algortihms and with certain scientific goals, to SDSS-V for optical spectroscopic observations as part of the Milky Way Mapper survey.}
{From our submitted candidates, a total of 4,946 unique targets were observed during SDSS-V DR20, of which 542 were identified as ACBs, with 538 likely CVs, and four being known X-ray binaries. Apart from the cartons that we submitted, we also identified CVs that were observed in other \erosita\ based cartons, specifically BHM SPIDERS, in which we identified an additional 49 systems that are likely CVs and which were not in our selection. We therefore identified 587 systems as CVs from the eROSITA based SDSS-V observations, 274 of which were previously known to the literature. We also attempted to sub-classify the CVs as dwarf-novae, novalikes, or magnetic systems based on the \erosita\ X-ray data, the emission line properties in the optical SDSS-V spectra, and other data in the public domain, such as optical light curves from surveys such as CRTS/ZTF and ATLAS, if available.}
{}

\keywords{binaries: close – novae, cataclysmic variables – X-rays: binaries}

\maketitle

\section{Introduction}

Throughout history, there have been records of stars varying in brightness \citep[][]{Xi_ancient_novae, warner2003cataclysmic, ancient_aus}, with a subset of those being referred to as novae, or new stars, that seem to suddenly appear and then fade away over timescales of days to months. We now know that these novae occur in binary systems known as cataclysmic variables (CVs). These are semi-detached binaries where a white dwarf (WD), which is the primary and more massive component, is accreting material from a Roche-lobe filling low-mass main sequence secondary component, which is also referred to as the donor or companion star \citep[for a comprehensive description of CVs, see][]{warner2003cataclysmic, hellier_CV_book, belloni+schreiber23-2}. Historically, CVs were found by their optical variability. However, with the advent of multi-wavelength astronomy, different means have been incorporated to try to identify potential CVs, as their behaviour across the electromagnetic spectrum became better known \citep[see for e.g.][for CV properties at different wavelengths]{cv_uv, warner2003cataclysmic, cv_ir, golden_age_cv_survey,  cv_radio, schwope_first_systematic}. 
Indeed, surveys at wavelengths other than optical have proven to find previously unknown CVs, and in particular, the \rosat\ survey \citep[][]{rosat_mission} showed this to be the case in the X-ray regime \citep[][]{rosat_cv}. In fact, CVs were even observed in the very first X-ray surveys, such as Uhuru \citep[][]{uhuru_survey} and HEAO-1 \citep[][]{HEAO-1}. In CVs, material from the Roche-lobe filling companion star flows through the inner Lagrangian point, L$_{1}$, in a stream towards the WD. However, accretion onto the WD is strongly dictated by the magnetic field strength of the WD and the instantaneous mass transfer rate, $\dot{\text{M}}$, and CVs can be broadly categorized as non-magnetic and magnetic (mCVs). The means by which the accreting material from the companion accretes onto the WD differs greatly between magnetic and non-magnetic CVs.

In non-magnetic systems, an accretion disk forms that may extend all the way to the surface of the WD. The region where the disk meets the surface of the WD is called the boundary layer (BL), where the matter in the disk graze the surface of the WD. In high accretion rate systems the BL is optically thick and produces soft X-ray emission \citep{BL_x-rays}. In low accretion rate non-magnetic systems, on the other hand, hard X-ray emission originates from an optically thin BL \citep[][]{optic_thin_accret_disk1, optic_thin_accret_disk2, optic_thin_accret_disk3}. In some low accretion rate systems, it is possible that the BL does not reach the surface of the WD, and instead it gets evaporated into a hot and diffuse corona, which gives rise to hard X-ray emission \citep[][]{CV_xray_corona}. 

In mCVs we find two distinct subclasses, polars and intermediate polars (IPs). Polars, or AM Her systems \citep[][]{polar, The_Polars, PolarCat}, have magnetic fields B $\gtrsim 10^7$ G, resulting in the magnetospheric radius extending beyond the circularization radius of the stream, and hence no accretion disk can form. Instead, material in the accretion stream directly attaches to the magnetic fields of the WD and flow along the field lines to the magnetic poles of the WD, where the material hits the surface in accretion columns \citep[][]{The_Polars, polar_accretion_column}. IPs, or DQ Her systems \citep[][]{IP_general, evolution_of_mCV} have weaker magnetic fields than polars, typically 10$^6 \lesssim B \lesssim 10^7$ G, and an accretion disk may form, but it is truncated at the magnetospheric radius of the WD, however, the truncation radius of the disk not only depends on the WD magnetic field, but on $\dot{\text{M}}$, as the ram pressure from higher $\dot{\text{M}}$ can decrease the WD magnetospheric radius, thus allowing a disk to form closer to the WD \citep[][]{IP_DN}. In IPs material from the disk follows the magnetic field lines in accretion curtains to the polar caps of the WD \citep[][]{IP_accretion_curtains}, and, as in the case of polars, X-ray emission likely originates from a shocked accretion column near the surface of the WD \citep[][]{X_ray_IP}. 

Regardless of the accretion process, X-rays in CVs are produced near the surface of the WD during the accretion process \citep[][]{Mukai_X_ray}, which enables these systems to be discovered by X-ray surveys, such as the extended ROentgen Survey with an Imaging Telescope Array \citep[\erosita,][]{eROSITA_site} on board the Spektrum-Roentgen-Gamma mission \citep[\SRG,][]{erosita_site2}. \erosita\ has provided the deepest X-ray survey of the sky, with the stack of the first three \erosita\ All Sky Surveys, eRASS:3, detecting close to 2 million X-ray sources down to a flux of F$_X$ $\sim 2.7 \times 10^{-14}$ erg s$^{-1}$ cm$^{-2}$ for point-like objects in the soft 0.5–2 keV band (Ramos-Ceja et al., accepted), which will include CVs. 
Given this sensitivity, we are able to detect all CVs with log(L$_\text{X} /\text{erg s$^{-1}$}) \gtrsim 29.9$ within 500 pc, which includes most known CVs. In fact, the eRASS:3 sensitivity will allow us to detect the most X-ray luminous IPs out to $\sim$8 kpc.

The primary interest of this paper is to identify all the CVs in the eROSITA eRASS1 (i.e. the first \erosita\ All Sky Survey) and eRASS:3 data sets, with particular emphasis on the latter as it provides greater sensitivity, as numerous questions regarding these systems remain unanswered, chief amongst which is their space density. The space density of CVs still remains unknown, as previous samples suffer from small number statistics, or are biased to the brighter systems. In particular, \citet[][]{schwope_flux_limited_samples} determined space densities from various X-ray flux limited samples and found values ranging from $\rho = (0.95^{+0.13}_{-0.06}) \times 10^{-6} \, \text{pc}^{-3}$  to $\rho = (3.56^{+0.58}_{-0.33}) \times 10^{-6} \, \text{pc}^{-3}$, for non-magnetic CVs and assuming a scale height of h = 260 pc. They also found a space density of $\rho = (2.8^{+3.7}_{-1.2}) \times 10^{-8} \, \text{pc}^{-3}$ for the magnetic IPs. All of these findings suffer from observational bias towards the X-ray brightest CVs. Polars were also excluded in these X-ray flux-limited samples. 
\citet[][]{pala}, on the other hand, report on a 150 pc volume-limited sample, and find a space density of $\rho = (4.8^{+0.6}_{-0.8}) \times 10^{-6} \, \text{pc}^{-3}$, with an assumed scale height of h = 280 pc. Even though this sample is believed to be $\sim$80\% complete, it suffers from low-number statistics, consisting of only 42 CVs. \citet[][]{inight_300pc} compiled a 300 pc volume-limited sample of CVs consisting of 151 systems (assumed to be $\sim$50\% complete), and determined a space density of $\rho = 2.23 \times 10^{-6} \, \text{pc}^{-3}$, assuming h = 280 pc. Therefore, it seems that there is a large discrepancy in these studies. 

Another avenue for addressing some of these discrepancies will be through the millihertz gravitational wave emission of short period CVs that may be detectable with the upcoming Laser Interferometer Space Antenna \citep[\lisa,][]{lisa}. Of the order of 10$^2$ CVs may be resolvable sources with \lisa\ \citep[][]{lisa_cvs}, with those clustering near the minimum period \citep[][]{pmin_cvs} providing stronger constraints on CV space density estimates. The benefit of resolving CVs with \lisa\ is that constraints to the space density estimates will not suffer from the typical electromagnetic biases. 

In addition to putting better constraints on the space density, having an even larger sample of CVs will help to understand various other outstanding questions in close binary evolution, including the strength of angular momentum loss due to magnetic braking in these systems \citep[][]{magnetic_braking_aml}, the orbital period distribution of non-magnetic and mCVs \citep[][]{schreiber_period_gap} and the origin of the WD magnetic field in mCVs \citep[][]{schreiber_magnetic_WDs}. It will also better constrain the contribution of CVs, in particular IPs, to the Galactic Ridge X-ray Emission \citep[GRXE,][]{GRXE}. 
However, the X-ray data from \erosita\ are not enough to determine whether an object is a CV. To conclusively determine that a system is a CV, an optical spectrum needs to be taken, which should reveal emission lines from an accretion disk or stream. 

We thus submitted $\sim$ 90,000 unique \erosita\ detected CV candidates to Sloan Digital Sky Survey V \citep[SDSS-V][]{SDSS-V_cite} to have optical spectra obtained as part of the Milky Way Mapper survey, of which $\sim$5,000 were observed, and will form part of SDSS DR20 \citep[][Griffith et al., in prep]{SDSS-V_cite}. It must be noted that the optical spectra of X-ray binaries (XRBs), especially those of low-mass X-ray binaries (LMXBs), are very similar to those of CVs, as they too show Balmer and \ion{He}{} emission lines due to accretion. These systems do not contain a white dwarf accretor, but instead host a neutron star or black hole as accretor, and hence are not CVs and, even though being accreting compact objects, are considered as contaminents in our study. Here we present the results of these observations.  

The paper is structured as follows, in Section \ref{sec:candidate_selection} we discuss the \erosita\ observations, as well as the three different strategies utilized to obtain the suspected \gaia\ optical counterpart to the X-ray source. In Section \ref{sec:SDSS_observations} we talk about the SDSS-V spectroscopic observations of our CV candidates, while in Section \ref{sec:results} we analyze the results of these observations. Lastly, in Section \ref{sec:discussion_conclusion} we discuss these findings, with appendices consisting of supplementary tables and figures to the main text.  


\section{Candidate selection} \label{sec:candidate_selection}

\subsection{\erosita\ observations} \label{subsec:eRSOITA_observations}

\erosita, orbiting in a wide halo-orbit about the Sun-Earth L2 point, scanned the entire sky in the 0.2-10 keV band, completing one scan every six months. Each completed scan is referred to as an eRASS (eROSITA All Sky Survey), thus the first  three completed scans are eRASS1, eRASS2, and eRASS3, respectively, and the combined stack of the first three scans is referred to as eRASS:3. Fourth and fifth scans were also undertaken, however the fifth was only completed to $\sim$40\%. The proprietary data rights for \erosita\ are split between Russia and Germany, with the German \erosita\ Collaboration (eROSITA\_DE) having data rights to the western Galactic hemisphere. eROSITA\_DE released the first calibrated catalogue of eRASS1, published in 2024 \citep[][]{merloni_erosita}. eROSITA\_DE also created a calibrated catalogue of eRASS:3, which is available to the collaboration and will be made public in the next \erosita\ data release, DR2 (Ramos-Ceja et al., accepted). While eRASS1 detected $\sim9.3 \times 10^5$ sources in the more sensitive 0.2-2.3 keV band at a flux limit of F$_X$ > $ 5 \times 10^{-14}$ erg cm$^{-2}$ s$^{-1}$ \citep[][]{merloni_erosita}, eRASS:3 increased this number to $\sim 2 \times 10^{6}$ sources, with F$_X$ $\sim 2.7 \times 10^{-14}$ erg cm$^{-2}$ s$^{-1}$ in the same band (Ramos-Ceja et al., accepted), that will be parallel to SDSS-DR20 (Griffith et al., in prep). Hence, stacking of the eRASS scans allows fainter sources to be detected, but having photons from various scans also allows the source location to be better determined, therefore increasing the likelihood of the correct optical counterpart to be selected for further studies. We defined three cartons \citep[for a detailed description of SDSS-V cartons, see ][]{SDSS_cartons} for SDSS-V spectroscopic follow-up observations by implemented three different approaches in an attempt to identify accreting compact binaries (ACBs), specifically CVs, in the eROSITA data, two of which were based on eRASS1, and one on eRASS:3. In Sections \ref{subsec:mwm_compact_Gen}--\ref{subsec:mwm_compact_BOSS} we describe the selection algorithms used to identify CV candidates that were submitted to the three respective cartons.

\subsection{MWM eROSITA compact Gen} \label{subsec:mwm_compact_Gen}

One of the approaches employed to find the optical counterpart to the X-ray source of the possible ACB was to use positional matching, i.e. selecting the nearest \gaia\ source to the eRASS1 X-ray position. For this, a positional corrected input catalogue was used, which contains the coordinates of 730,174 eRASS1 X-ray sources \citep[for details on the positional correction applied to eRASS1, see][]{merloni_erosita}. The following cuts to the input catalogue were made in \texttt{TOPCAT} \citep[][]{topcat}. To ensure only point-like sources were selected, we set \texttt{EXT} < 6", as extended sources in this parameter typically have values ranging between 10" - 60". We next imposed a cut in the detection likelihood of the sources. The criteria of the cut were \texttt{DET\_LIKE\_1} >= 8 || \texttt{DET\_LIKE\_2} >= 8 || \texttt{DET\_LIKE\_3} >= 8, with \texttt{DET\_LIKE\_1}, \texttt{DET\_LIKE\_2}, and \texttt{DET\_LIKE\_3} being the detection likelihood of a source in the 0.2-0.6 keV, 0.6-2.3 keV, and 2.3-5.0 keV bands, respectively. This ensures that we select all targets of interest, including those that show purely soft, or purely hard X-ray emission. 511,823 sources remained after imposing these two initial cuts.

The next set of cuts were based on \gaia\ DR2 properties, or a combination of \gaia\ and \erosita\ properties. The first of these cuts were based on the \gaia\ position of a source in relation to the X-ray position. We selected those objects that had an uncertainty measure in their position \texttt{radec\_err} > 0, and \texttt{angdist}/\texttt{radec\_err} < 2.1, which is the angular distance of the match over the uncertainty of the position. The next cut was based on the ratio of the X-ray to optical flux of the object. This was done in an attempt to remove stars that are coronal emitters. We use the same definition of log(F$_\text{X}$/F$_\text{opt}$) as was used by \citet[][]{schwope_first_systematic}, i.e. log(F$_\text{X}$/F$_\text{opt}$) = log(F$_\text{X}$) + G$_\text{mag}$/2.5 + 4.86, with F$_\text{X}$ being the flux as measured in the 0.2-2.3 keV band, and G$_\text{mag}$ being the \gaia\ DR2 photometric mean magnitude in the G-band\footnote{\url{https://www.cosmos.esa.int/web/gaia/iow_20180316}}. The constant 4.86 was determined specifically for the \gaia\ G-band by integrating a folded flat spectrum and normalizing it to mag$_\text{AB}$ = 0. By setting log(F$_\text{X}$/F$_\text{opt}$) > -2.7, we were left with a, presumably, non-stellar subset of 430,094 objects. To ensure that we only selected Galactic sources, a cut was made based on the parallax and proper motion of the targets. The expression used was to not include those systems that satisfy \texttt{[(logpmdpm < 0.301 AND (parallax / parallax\_error) > -1.4995 $\times$ logpmdpm - 4.05 AND (parallax / parallax\_error) < 1.4995 $\times$ logpmdpm + 4.05)]} or those who satisfy \texttt{[(logpmdpm - 0.301) $\times$ (logpmdpm - 0.301) / (0.39794 $\times$ 0.39794) + (parallax / parallax\_error) $\times$ (parallax / parallax\_error) / (4.5 $\times$ 4.5) <= 1)]}. Here \texttt{pm} is the proper motion and \texttt{logpmdpm} $\equiv$ log(pm / pm\_error). Applying this cut significantly reduced the number of extragalactic objects. The final cut was made on the distance to the target. We chose to only include those targets that have \gaia\ DR2 distances \citep[][]{gaia_dist_bailer_jones} that lie within 3 kpc, i.e. \texttt{rest} < 3000. Applying all the above-mentioned cuts, our final target selection consists of 78,252 targets. As the resulting carton, mwm\_erosita\_compact\_gen, was based on eRASS1, it formed part of the SDSS-V v0.5 targeting cartons, and hence was only observed at APO.

\subsection{MWM eROSITA compact Var} \label{subsec:mwm_compact_Var}

Another method to find the optical counterpart to the X-ray source was to assign the optically most variable source within a radius of 30" of the eRASS1 coordinates as the counterpart. This was done by following all the same steps as in Section \ref{subsec:mwm_compact_Gen}, but also now introducing an optical variability cut. The proxy for variability observed in the \gaia\ data that we employed is the same as was used by \citet[][]{varproxy}, and is given as \texttt{varproxy = log$\sqrt{G_\text{n\_obs}/G_\text{FOE}}$}, where \texttt{$G_\text{n\_obs}$} is the number of observations by \gaia\ of the source in the G-band, and \texttt{$G_\text{FOE} = \langle G_\text{flux}\rangle / \langle G_\text{flux\_err}\rangle$}, i.e. the mean \gaia\ G-band flux divided by the error in the flux measurement. We also implemented the same relation for non-variability as used by \citet[][]{varproxy}, which is defined as \texttt{nonvar $= -\, 0.00018371431841165834 \times \langle G_\text{mag}\rangle ^3 + 0.0229179363443418 \times \langle G_\text{mag}\rangle ^2 -0.39976106099662695 \times \langle G_\text{mag}\rangle - 0.7999850524873042$}. See \citet[][]{varproxy} Figure 1 for a visual representation of this \texttt{nonvar} function. The sigma of the variability is given by \texttt{$\sigma_\text{var} = 0.0005839694693204828 \times \langle G_\text{mag}\rangle ^3 + 0.02926085703984754 \times \langle G_{mag}\rangle ^2 + 0.48185670698065436 \times \langle G_\text{mag}\rangle - 2.554121286005859$}. Lastly, the condition for a source to be considered as being variable was defined as \texttt{var > nonvar $\times$ 3$\sigma_\text{var}$}. Applying this additional cut, together with those outlined in Section \ref{subsec:mwm_compact_Gen}, a total of 17,058 targets were identified as the likely \gaia\ optical counterparts to the eRASS1 sources. The resulting carton, mwm\_erosita\_compact\_var, also formed part of the SDSS-V v0.5 cartons, and hence was only observed at APO.        

\subsection{MWM eROSITA compact BOSS} \label{subsec:mwm_compact_BOSS}

\begin{figure}[] 
     \centering
     \begin{subfigure}[b]{0.5\textwidth}
         \centering
        \includegraphics[width=\textwidth]{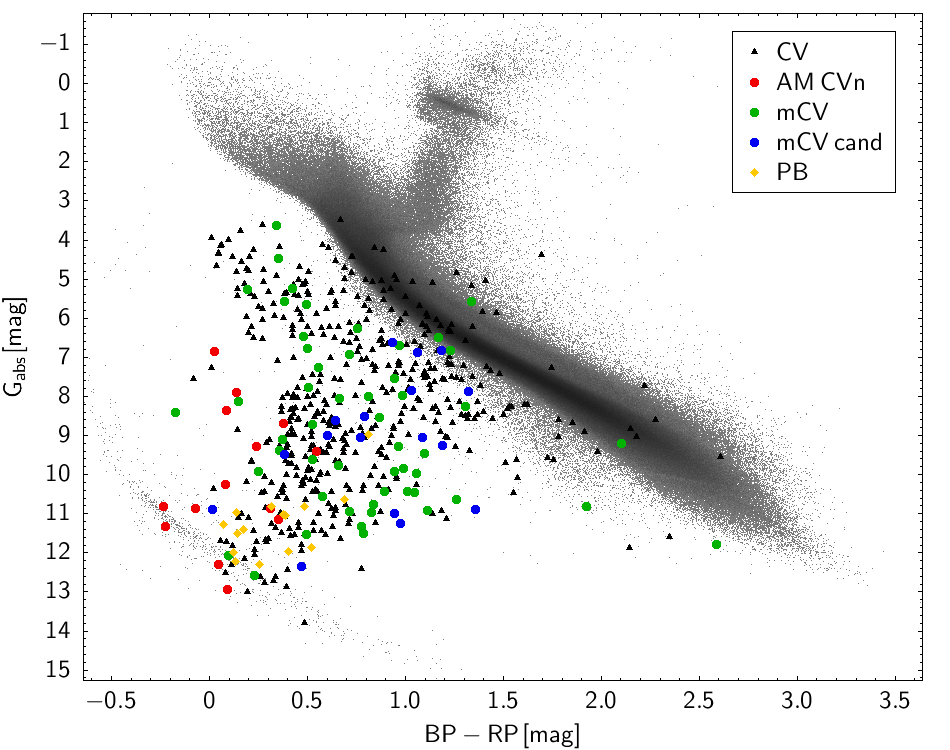}
        \caption[DR20_CM]{Colour-magnitude}
        \label{fig:CM_selection}
     \end{subfigure}
    \hfill
     \begin{subfigure}[b]{0.5\textwidth}
         \centering
        \includegraphics[width=\textwidth]{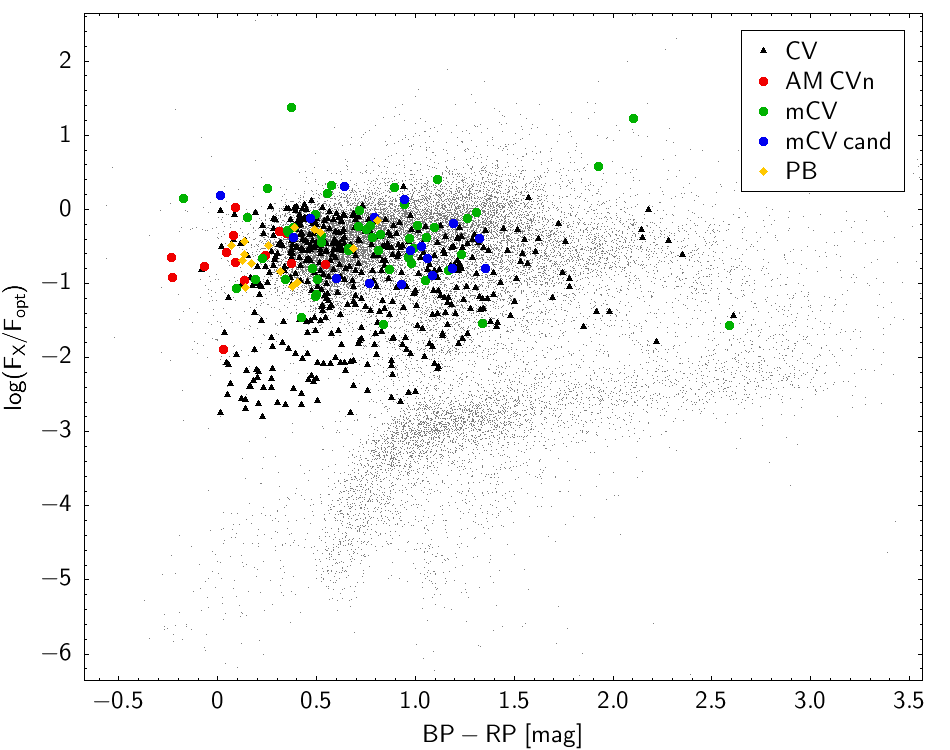}
        \caption[DR20_CC]{Colour-colour}
        \label{fig:CC_selection}
     \end{subfigure}

\caption{Diagnostic plots, identifying the location of the CVs reported in this study (see Section \ref{sec:results}) in the respective diagnostic spaces. The background in (a) comprises \gaia\ DR3 sources with well established distances, while the background in (b) consists of \gaia\ DR3 sources with known X-ray fluxes. We identify our CVs with black triangles, with specific subtypes highlighted. In red we have the AM CVns, green are the known mCVs, while in blue we have the mCV candidates identified in this work, and in orange diamonds we have the period bouncers (PB).}
        \label{fig:diagnsotic_plots}
\end{figure}

The third approach we took was to incorporate machine learning (ML) in the likely CV candidate selection. First, using the \gaia\ EDR3\footnote{\url{https://www.cosmos.esa.int/web/gaia/earlydr3}} optical photometric and eRASS1 X-ray 0.2-2.3 keV band properties, we identified where these systems lie in parameter spaces, such as log(F$_\text{X}$/F$_\text{opt}$) against BP - RP magnitude (colour-colour), as well as G-band absolute magnitude against BP - RP magnitude (colour-magnitude). As an example, Figure \ref{fig:diagnsotic_plots} gives the location of the CVs reported in this study (see Section \ref{sec:results}), and shows that these systems generally occupy very distinct regions in the respective diagnostic spaces. One property of CVs that is immediately apparent when studying these plots, is that they are typically very blue objects, with the vast majority having BP-RP < 1.5 mag, which is a result of the accretion disk, or accretion process for mCVs, in these systems. 

CVs are found almost exclusively between the main-sequence track and the WD branch (Figure \ref{fig:CM_selection}), with those systems lying on the WD branch usually being very low-accreting systems, including the so-called period bouncers (PBs), which contain a brown dwarf, or other low-mass degenerate object, as companion \citep[see also Hernandez-Diaz et al. 2026 (submitted), who are reporting PBs observed by SDSS-V]{period_bounce}. Figure \ref{fig:CC_selection}, further shows that CVs are found on the active galactic nuclei (AGN) branch, and not the stellar branch, when looking at their optical to X-ray fluxes, thus further distinguishing them from coronal emitting stars. Using the location in these diagnostic spaces as criteria, together with the proper motion and distance obtained from \gaia\ (with the criterion that the sources must be Galactic), a training sample of known CVs were used to train a Bayesian based ML algorithm \citep[incorporating NWAY\footnote{\url{https://github.com/JohannesBuchner/nway/}},][]{NWAY} to identify the correct \gaia\ optical counterpart to the eRASS:3 X-ray sources. 

We trained a Random Forest (RF) classifier to distinguish CVs from other sources (such as AGN, coronal emitters etc.), using select X-ray (eRASS1) and optical \gaia\ EDR3 features,
namely, \texttt{parallax}, \texttt{parallax\_over\_error}, \texttt{phot\_g\_mean\_mag}, \texttt{phot\_g\_mean\_flux\_over\_error}, \texttt{phot\_bp\_mean\_mag}, \texttt{phot\_rp\_mean\_mag}, \texttt{varproxy}, \texttt{logpmddm}, \texttt{phot\_bp\_mean\_flux\_over\_error}, \texttt{bp-rp}, and \texttt{logfxfopt}. A dataset consisting of 624 known-CVs and $\sim$90,000 secure non-CVs was split 70/30 to train and test the model, respectively.
The RF learns to map the tendency of the sources in this multidimensional parameter space into a CV-probability (P$_\text{cv}$). To select CV candidates from eRASS:3, we first cross-matched the eRASS:3 1-band catalogue with \citet[][]{bailer-jones_correct} distances and retrieved all optical sources within a 30 arcsec radius of the X-ray position. For each of the X-ray-optical association we determine P$_\text{cv}$ using the trained RF-model. Subsequently, we use the P$_\text{cv}$ as prior-probability for the Bayesian cross-match tool NWAY to evaluate the optical counterparts. NWAY combines the prior-probability (here P$_\text{cv}$) with distance-based probability to compute two posterior probabilities, P$_\text{any}$ and P$_\text{i}$. P$_\text{any}$ gives the probability that there exists a true optical counterpart to the X-ray source and P$_\text{i}$ gives the relative probability between all potential counterparts. We restrict ourselves to counterparts with a P$_\text{any}$ $\geq$ 0.2 and P$_\text{cv}$ $\geq$ 0.5, in case of multiple counterparts satisfying these criteria we keep the first two with the highest relative probability i.e. P$_\text{i}$. We also remove duplicated \gaia\ counterparts and only keep the association with the highest P$_\text{cv}$. Additionally, we enforce a hard-cut on the \gaia\ colour and keep only those with BP-RP < 2, we also remove all candidates that occupy the stellar branch in the colour-colour diagram. The 11,113 candidates that remain were selected as targets for SDSS-V.

As this carton, mwm\_erosita\_compact\_boss, was based on eRASS:3 and \gaia\ DR3, it formed part of the SDSS-V v1.0 targeting cartons, and was thus observed at both APO and LCO.

\section{SDSS-V observations} \label{sec:SDSS_observations}

The Sloan Digital Sky Survey V (SDSS-V) started observations in 2021 \citep[][]{SDSS-V_cite}. It is an optical-to-infrared, multi-epoch, all-sky spectroscopic survey, which utilizes the 2.5m f/7.5 du Pont Telescope \citep[][]{duPont_telescope} at the Las Campanas Observatory (LCO), Chile, and the 2.5m f/5 Sloan Foundation Telescope \citep[][]{SDSS_telescope} at the Apache Point Observatory (APO), in The United States of America. Each telescope has an identical multi-object spectroscopy (MOS)-spectrograph, BOSS \citep[][]{BOSS_spectrograph}, which was originally built for the SDSS-III Baryon Oscillation Spectroscopic Survey \citep[][]{BOSS_survey}. During the first year of SDSS-V observations, the BOSS spectrographs used a plate system for its multi-object spectroscopic observations. This was later changed to a focal plane system (FPS)\footnote{\url{https://www.sdss.org/instruments/fps/}}, which consists of 500 fibres, thus allowing 500 spectra to be taken simultaneously. Science observations using the FPS started in early 2022 at APO, while at LCO, the science operations started at the end of 2022. As the focal length of the two telescopes are not the same, the fibre diameter is 2" at APO, while at LCO it is 1.5". The spectral range covered by BOSS is 3600-10,400\AA, however this is split in two channels, a blue and a red, with the split at $\sim$6000\AA. The blue channel has a resolution of R=1560-2270, while the red has a resolution R=1850-2650. The main science goals of SDSS-V are encapsulated in three surveys, Black Hole Mapper (BHM), Milky Way Mapper (MWM), and Local Volume Mapper (LVM). For an extensive description of each survey, see \citet[][]{SDSS-V_cite}. 

All three of our cartons (see Sections \ref{subsec:mwm_compact_Gen}–\ref{subsec:mwm_compact_BOSS}) form part of the MWM survey, as the focus of this survey is to study Galactic systems. However, we expect that CVs will also be observed in the BHM survey. BHM is set to be the first comprehensive spectroscopic follow-up of X-ray sources observed by \erosita, as the vast majority of these sources are expected to be AGN. The Spectroscopic Identification of eROSITA Sources (SPIDERS) programme is therefore conducting optical spectroscopic follow-up of X-ray sources observed in the western Galactic hemisphere, i.e. that available to the eROSITA\_DE Consortium. SPIDERS is set to provide identification spectra and redshifts for $\sim$350,000 eRASS:3 sources, with these sources having F$_{0.5-2\,\text{keV}} \gtrsim 2\times 10^{-14}$ erg s$^{-1}$ cm$^{-2}$. Even though the vast majority of objects observed in BHM-SPIDERS will be AGN and QSOs, the survey will inevitably observe other X-ray sources as well, including CVs \citep[][]{SDSS-V_cite}. See Merloni et al., in prep, who will report on the results of the SPIDERS programme.     

The observations that form part of SDSS DR20 concluded on MJD 60708 (2 February 2025), and all of our results are based on processing version v6\_2\_1 of the idlspec2d pipeline that was used to reduce the BOSS spectra for SDSS DR20.


\section{Results} \label{sec:results}

An optical spectrum is one of the most telling indicators of whether an object is a CV or not, as it would reveal Balmer, and possible \ion{He}{i} and \ion{He}{ii} emission lines, indicative of the accretion process occurring in these systems. However, having other optical and X-ray data at hand can aid in the subclassification of a system, and we therefore utilised these in our analysis of each SDSS-V observation, in conjunction with the spectrum obtained. In particular, the X-ray hardness ratio (HR) in adjacent bands, where HR = (H-S)/(H+S), with H being the hard band and S the soft, can be very telling of whether a system is magnetic, as mCVs typically have a strong soft X-ray component \citep[][]{polar_soft_excess, IP_soft_excess, Mukai_X_ray}. Thus, we define two hardness ratios, with HR P12 being defined with S = 0.2-0.5 keV, H = 0.5-1.0 keV, and, HR P23 with S = 0.5-1.0 keV, and H = 1.0-2.0 keV. 

Figure \ref{fig:V1309_Ori_diagnostic} gives an overview of the diagnostic tools we used in our screening, illustrating this for the known eclipsing polar V1309 Orionis (V1309 Ori), which was re-observed during SDSS-V. Figure \ref{fig:V1309_Ori_SDSS_V_spectrum} is the SDSS-V spectrum obtained, and clearly shows prominent Balmer and Paschen emission lines, as well as \ion{He}{i} and \ion{He}{ii} emission lines, with \ion{He}{ii} $\lambda$4686 being particularly strong, together with the Bowen Blend. In Figure \ref{fig:V1309_Ori_finding_chart} we have the finding chart, indicating the eRASS:3 positioning, as well as the SDSS-V fibre position, thus confirming that the correct target was observed. Figures \ref{fig:V1309_Ori_colour_mag} and \ref{fig:V1309_Ori_colour_colour} show the colour-magnitude and colour-colour diagrams, respectively. Lastly, Figures \ref{fig:V1309_Ori_ATLAS_LC} and \ref{fig:V1309_Ori_ZTF} show the ATLAS and CRTS/ZTF light curves, respectively. While the ATLAS data do show variability, the ZTF light curve clearly shows the $\sim$40-minute eclipses \citep[][]{V1309_Ori, PolarCat}, which is difficult to see in the ATLAS light curve, thus illustrating the benefit of having multiple datasets to utilize. 

In addition to these plots, we also have the flux measurement in the 0.2–2.3 keV band, from which we can determine the X-ray luminosity of every target, using the \gaia\ DR3 distances, which, for V1309 Ori is log(L$_\text{X}/\text{erg s$^{-1}$}$) = 32.15. Lastly, for the HR we find HR P12 = -0.85 $\pm$ 0.02, and HR P23 = -0.64 $\pm$ 0.08 for V1309 Ori, indicating that the X-ray emission is almost completely dominated by soft X-rays, and hence confirming the magnetic nature of the system. Thus, using the optical spectrum, together with the additional optical photometric and X-ray data, we are able to not only determine whether a system is a CV or not, but we can, in many cases, determine the subclass to which the system belongs.     

As the primary goal of these SDSS-V observations was to establish whether our candidates are indeed CVs, the observing strategy was to obtain one, or possibly two, spectra of each target, as this would be sufficient to prove or disprove the CV nature of the system. However, targets are not necessarily exclusive to a single carton, and many of our targets were also in other cartons, 
and hence obtaining spectra on numerous epochs. This proved very fortuitous, as some spectra obtained turned out to be very noisy, making spectral line identification a challenge, and an accurate determination of a system could only be done as a result of additional spectra taken on different epochs (see Figure \ref{fig:appen_J135436_positioning} in the appendix as an example). However, 
there were also candidates that had only one spectrum which was very noisy, possibly due to external influences such as bad seeing or other factors, and hence a positive identification of the nature of the source could not be established. 

Here we report on those systems for which a positive ACB identification was determined with a high degree of certainty based on the observed spectrum, and aided by the auxiliary information we have on each target, as discussed above. Table \ref{tab:cv-submit_observed} summarizes the results of the screening done for the spectra obtained up to 2 February 2025 (MJD 60708, the cutoff date for SDSS-V DR20), for the three respective cartons, and shows that only a small fraction of our targets were observed, together with the success rate of ACBs observed per carton. These ACBs consist of CVs, including the double-degenerate AM CVn systems \citep[][]{AM_CVns_nelemans}, as well as X-ray binaries (XRBs). AM CVns consist of a WD accreting from another degenerate object and hence we include them in our classification as a CV, although they do form a distinct class of objects by themselves (Section \ref{subsec:AM_CVns}). XRBs, on the other hand, contain an accreting neutron star or black hole, and hence are not considered as CVs, and are removed at a later stage from our results (see Section \ref{subsec:crossmatch}).

\subsection{MWM cartons} \label{subsec:MWM_cartons_results}

\begin{table}
\centering 
\caption{SDSS-V \erosita\ accreting compact binary candidate screening} 
\label{tab:cv-submit_observed}
\begin{threeparttable}
\begin{tabular}{cccccc}
\hline
\hline
Carton  & eRASS & Targets   & Observed  & ACB  & Success \%  \\
\hline
1       &  1   &  78,252   & 3,281     & 189   & 5.8   \\
2       &  1   &  17,058   & 1,850     & 199   & 10.8  \\
3       &  :3  &  11,113   & 1,658     & 523   & 31.5  \\
\hline
Unique  &      &  89,337   & 4,946     & 542   & 11.0  \\
	\hline
	\hline
\end{tabular}
\begin{tablenotes}
    \item Carton 1: mwm\_erosita\_compact\_gen 
    \item Carton 2: mwm\_erosita\_compact\_var
    \item Carton 3: mwm\_erosita\_compact\_boss
    \end{tablenotes}
\end{threeparttable}
\end{table}   

There was significant overlap of ACB candidates amongst the three cartons, hence, although a total of 106,423 targets were submitted amongst the three cartons, only 89,337 were unique, of which 4,946 were observed. Of those observed targets, we found that 542 were indeed ACBs, corresponding to an overall success rate of 11.0\%. However, looking at each carton individually tells a different story. The mwm\_erosita\_compact\_gen carton had by far the most targets submitted, however only 4.2\% were observed, of which only 5.8\% were ACBs. In the mwm\_erosita\_compact\_var carton 10.8\% of the targets were observed, of which 10.8\% were ACBs. However, the mwm\_erosita\_compact\_boss carton had the least number of candidates, and the least total number of observed targets (with 14.9\%), but had the highest number of ACBs, with a success rate of 31.5\%. 

It is interesting to note that the mwm\_erosita\_compact\_gen carton contained 16 systems that were not in the mwm\_erosita\_compact\_boss carton, while the mwm\_erosita\_compact\_var carton contained 17 that were not in the mwm\_erosita\_compact\_boss carton. There was significant overlap between these cartons, however, a total of 19 unique systems were amongst the two cartons based on eRASS1 (i.e. mwm\_erosita\_compact\_gen and mwm\_erosita\_compact\_var), which were not amongst the eRASS:3 based ML carton, i.e. mwm\_erosita\_compact\_boss. This discrepancy can be explained for four of the systems, since three of them have BP - RP > 2 mag, while one does not have a BP - RP value, and hence would not have made the colour cut imposed in the mwm\_erosita\_compact\_boss carton selection criteria. However, the explanation as to why the other 15 systems were not chosen as ACB candidates in the ML approach is unclear at present. However, it is interesting to note that 14 of the 19 systems, including the three systems with BP - RP > 2 mag, were identified in the various BHM SPIDERS cartons, and hence the ML techniques employed in mwm\_erosita\_compact\_boss, simply identified the wrong \gaia\ optical counterpart as the likely ACB candidate. It is beyond the scope of this paper to fully investigate why a different optical counterpart was chosen between the two ML algorithms.

Given this, there are still seven systems identified in eRASS1 as ACB candidates that did not appear in the eRASS:3 ML candidate selection. To investigate the possible reason for this, we ran a positional search in \texttt{TopCat}, and found no eRASS:3 entries to within 30" of those seven eRASS1 targets. It therefore seems likely that these seven systems fell below the detection likelihood, \texttt{DetML}, implemented in the eRASS:3 source detection, and hence why there are no entries for them in eRASS:3. This should not be too surprising, as CVs are transient systems, and can have high- and low-states. We therefore conclude that these seven systems were likely in a high state during the eRASS1 observations, and were observed during low states in the subsequent eRASS observations.

\subsection{BHM SPIDERS carton} \label{subsec:spiders}

Machine learning \citep[][]{NWAY} was also incorporated for the optical counterpart identification of the SPIDERS targets. These observations formed part of the SDSS-V BHM programme and did not form part of the cartons we submitted under the MWM programme (Sections \ref{subsec:mwm_compact_Gen} -- \ref{subsec:mwm_compact_BOSS}). Visual inspection of these observations were conducted, and will be reported in Merloni et al. (in prep), who also details the target identification criteria. 

The primary goal of SPIDERS is to perform a maximally complete AGN survey. However, optical counterparts to all eROSITA point-like X-ray sources were also selected for spectroscopic follow-up. In the context of the current paper, Galactic X-ray emitters are important, but the automatic pipeline classification often fails for the unusual spectra of ACBs. 
Therefore, in the context of the current paper, careful visual inspection was performed i) for all spectra that were labelled as STAR by the SDSS-V spectroscopic pipeline, and ii) on those that were labelled as likely stars, i.e. likely Galactic X-ray sources irrespective of their true nature, after a first screening run. 

The first screening was done with the main purpose to distinguish between certain extragalactic objects and distinguish them from Galactic sources. Such sources were screened a second or even a third time to efficiently filter for ACBs. This procedure eventually revealed a total of 198 ACBs, which include six AM CVn systems. Amongst these 198 systems, it was found that 149 overlapped with targets that were also identified in the MWM cartons we defined, while 49 were unique to the SPIDERS cartons. 
   
While 46 of these targets were identified as eRASS:3 sources, there were three systems that were not amongst the candidates from eRASS:3, and instead they were observed under the bhm\_spiders\_agn-efeds\_0.1.0 carton, i.e. AGN candidates identified in the eROSITA Final Equatorial Depth Survey \citep[eFEDS,][]{eFEDS}.
Of these 49 systems, we also found 20 that were observed under bhm\_spiders\_agn\_lsdr10, i.e. the optical counterpart matching was done using Legacy Survey DR10 \citep[][]{legacy_survey}. Two of these systems were also in the abovementioned eFEDS carton. However, to determine whether a \gaia\ DR3 counterpart exists for these systems, a crossmatch to \gaia\ EDR3 was done with a search radius of 2", within \texttt{Topcat} to the fibre positioning coordinates of the SDSS-V observation. This search result, however, did not deliver any matches.

Thus, we therefore identified an additional 49 ACBs that were not amongst our targets, and including these, resulted in 591 ACBs observed by SDSS-V from \erosita\ based candidates.

\subsection{Deriving the CV sample from the ACBs} \label{subsec:crossmatch}

We identified all known XRBs in our sample of ACBs by running a crossmatch within \texttt{TopCat} with the LMXB catalogue published by \citet[][]{LMXB_cat}, as well as the high-mass X-ray binary (HMXB) catalogue by \citet[][]{HMXB_cat}. While no HMXBs were in our 591 ACBs, we did, however, find four LMXBs. These systems were V616 Mon, SWIFT J0428.2-6704A, IGR J14298-6715, and PSR J1023+0038, the first discovered transitional millisecond pulsar \citep[tMSP,][]{first_tmsp,tMSP}.

Thus, after removing those four LMXBs, we are left with 587 systems which we regard, with a high degree of certainty, as CVs from our observed candidates. From literature searches of these 587 CVs, 
using existing CV catalogues \citep[][]{ritter_kolb, schwope_first_systematic, inight_2025, PolarCat,tess_cvs}, we found that 274 are already known as CVs, implying that many of our systems might be new to the literature. In addition to these 274 known systems, we also found 114 systems that only had entries in The International Variable Star Index (VSX)\footnote{\url{https://vsx.aavso.org/}} of the American Association of Variable Star Observers (AAVSO), with no other references.   

The literature search also revealed that 51 of the systems are known mCVs, with 13 subclassified as IPs, and 36 as polars, while two mCVs do not have a subclassification. We also found 11 known AM CVn systems amongst our 587 CVs during our literature search. A total of 278 DN were identified, including those with only a VSX entry, as well as 30 NL systems, and three systems classified as novae. Hence, we did not find records in any of the CV catalogues, and other references, for 199 of the 587 CVs, and thus these systems could be new discoveries. We did, however, make tentative classifications for 22 of these 199 systems, amongst which are four AM CVns, while the rest are mCVs, including one low accretion rate polar \citep[LARP,][]{LARPs, PolarCat}. In Section \ref{subsubsec:LARP} we discuss the LARP we identified, while in Section \ref{subsec:AM_CVns} we discuss the AM CVn systems we identified. In Section \ref{subsec:catalogue} we discuss the criteria we applied to identify the mCVs not listed in the literature.   

\subsection{New CVs within d < 150 pc}

Amongst the newly identified CVs, we find two systems that lie within 150 pc. The first system in question is 3eRASS J023101.9$-$585844, which is a LARP, while the second, 3eRASS J033833.9$-$332755, is a PB candidate. Further details of these two systems can be found in Appendices \ref{subsubsec:LARP} and \ref{subsubsec:new_PB}, respectively. Identifying new CVs within 150 pc is of special interest, as \citet[][]{pala} found 42 CVs within this volume, 22 of them in the western Galactic hemisphere, and they concluded that their sample contained $\sim$80\% of all the CVs within this volume. The number of CVs within this volume was increased when \citet[][]{rodriquez_new_150pc} found another two CVs within 150 pc. With the addition of the two systems we found in our study, the 150 pc volume-limited CV sample inches ever closer to being complete.

\subsection{Early-type donor CVs}

As CVs evolve from long to short orbital periods, this implies that, in order to fill its Roche-lobe, the spectral class of the companion should be different for long and short orbital period systems \citep[see][for a review of the orbital period to CV companion relation]{CV_companion_spectral_class}. The companion in the vast majority of CVs is an M-dwarf star, however, the long period systems typically contain a main-sequence companion of spectral class G or K. One defining feature in the spectrum of these stars is the presence of Mg absorption at $\lambda\sim$5170\AA. However, it is possible that the early-type donor could also be nuclearly evolved, and not just a typical main-sequence star, as shown by \citet[][]{evolved_CV} for another CV that was observed by SDSS. 

We identified 14 early-type donor systems, with the spectrum of one, 3eRASS J063026.9-652950, shown in Figure \ref{fig:appen_063027.19_G_K_companion}, together with a template spectrum of a K5 dwarf star \citep[adapted from][]{spectra_librarry}. The Mg absorption in both spectra is immediately apparent, as well as the Na D absorption at $\lambda\sim5895$\AA, which is another defining spectral feature in late-type stellar spectra \citep[see for example][]{Na_D}, although this should not be confused with possible interstellar absorption \citep[][]{ISM_Na_D}. In Figure \ref{fig:appen_063027.19_G_K_companion} we also see a rising blue continuum, which is due to a combination of, primarily, the emission from the accretion disk, as well as the WD, while the prominent Balmer emission lines is further evidence of accretion occurring in this system. Table \ref{tab:early_type_donors} in the Appendix contains the IAU names of these 14 systems, which can be used to identity them in the catalogue accompanying this paper (this is described in Section \ref{subsec:catalogue}). 

\subsection{AM CVn systems} \label{subsec:AM_CVns}

We also identified 14 AM CVn systems, four of which did not appear in our literature searches, and thus could be new discoveries (see Hernández-Díaz et al., submitted, who reports on more AM CVn systems observed beyond MJD 60708, i.e. the SDSS DR20 observation cutoff date). These systems have very short orbital periods, typically 5–65 minutes, and are also known as double-degenerate systems as the companion is also a degenerate object such as a WD, or a semi-degenerate He-rich star \citep[][]{AM_CVns}. Thus, the defining spectral feature in AM CVns is helium emission lines, in combination with the complete lack of hydrogen emission.

That being said, we also found one system, 3eRASS\,J080449.4+161625, that shows weak emission that could be H$\alpha$ and H$\beta$ Balmer lines, as is seen in Figure \ref{fig:appen_080449_Am_CVn}, however \citet[][]{AM_CVn_with_additional_He} suggest that the emission is from \ion{He}{ii} $\lambda$6559 and \ion{He}{ii} $\lambda$4860 Pickering series instead. 

If the emission was due to hydrogen, the presence of residual hydrogen is unusual for a classical AM\,CVn star and may indicate an evolved donor that has not yet lost all of its hydrogen envelope \citep[][]{AM_CVn_evolved_donor_cv}. Similar systems have been reported previously and are often interpreted as transitional objects linking hydrogen-rich cataclysmic variables to the helium-dominated AM CVn population 
\citep[e.g.][]{burdgeetal22-2,greenetal25-1}. 
The existence of such transitional objects as well as the relatively large number of 
CVs with evolved donors \citep{el-badryetal21-1}, support the idea that AM\,CVn could form from CVs \citep{belloni+schreiber23-1}.

\subsection{Contaminants}

Several different object classes appeared as contaminants in our observed candidates. Table \ref{tab:contaminants} lists those object classes that were of most interest, as they might reveal potentially overlooked selection criteria. The object classes that we list are M-stars, WDs, detached WD and M-star binaries, as well as  young stellar objects (YSOs). We also group together all extragalactic sources, i.e. galaxies (predominantly starburst galaxies), active galactic nuclei (AGN), and quasars (QSOs), and we make no attempt to further distinguish between the extragalactic objects.

We also did not attempt to sub-classify the WDs, however, we did identify 11 that are He WDs (known as DB WDs). Many of the WDs, however, did show weak Balmer emission cores in the absorption lines, with no indication of a companion in the spectrum. Typically, Balmer emission lines would be an indication of possible accretion occurring in a system, however, a subclass of isolated magnetic WDs, known as DAe WDs, exhibit weak Balmer emission cores, while not showing any Zeeman-line splitting \citep[][]{WD_balmer_emission}, which is usually seen in isolated magnetic WDs. However, we did not determine the magnetic fields of these systems, and their classification as DAe WDs thus needs to be confirmed by additional observations. 

There were also contaminants of normal main-sequence stars, apart from the M-stars, of most spectral classes from B to K. Some of the A- and B-stars showed very narrow Balmer emission cores in the absorption lines, which could be due to rapid rotation in these stars, or indicate that these are young pre-main sequence Herbig Ae/Be stars, which are still surrounded by a circumstellar debris disk \citep[][]{herbig_stars}. It is also likely that some of the targets we classified as YSOs could in fact be T Tauri stars \citep[][]{t_tauri_stars}. The remaining spectra were either very noisy and difficult to classify, or the fibre position was not on target and subsequently a sky spectrum was taken. 

\begin{figure}[h!]
        \centering
        \includegraphics[width = \columnwidth]{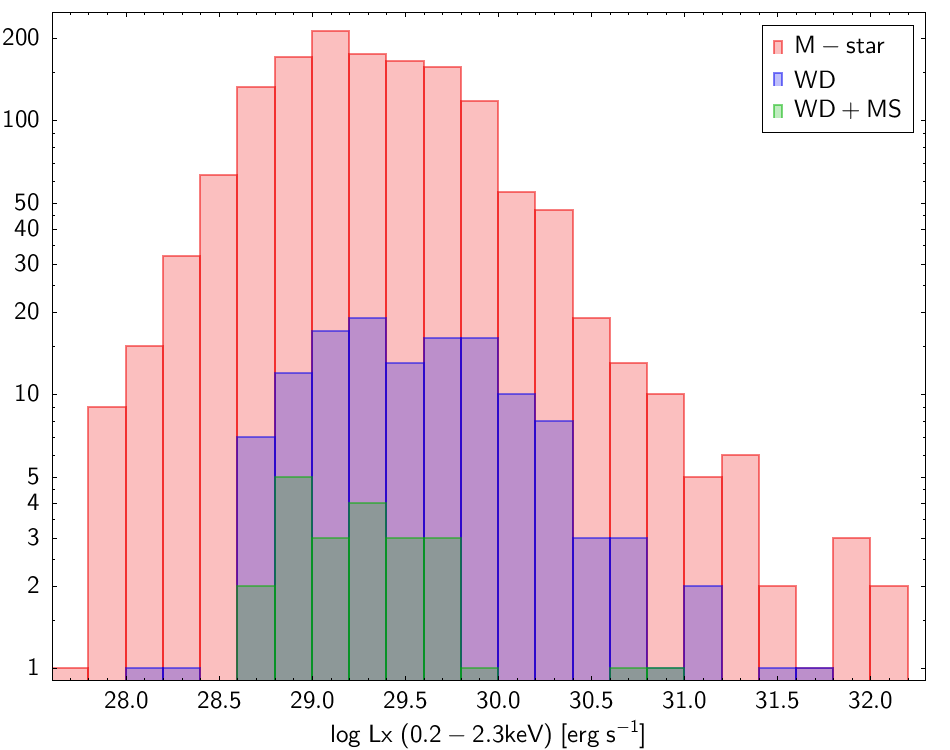}
        \caption[histogram contaminants Lx]{Histogram of log(L$_\text{X}$) in the 0.2 - 2.3 keV range of the main contaminants that could harbour quiescent CVs, with M-stars in red, WDs in blue, and WD + main sequence binaries in green. Note the logarithmic scale of the y-axis.}
        \label{fig:hist_Lx_contaminants}
\end{figure}

It is possible that quiescent CVs could be hiding amongst the contaminants, particularly among the M-star, WD, and WD + main sequence binaries. A weakly accreting system could have been accreting during the \erosita\ observations, hence being detected as an X-ray source, but have since gone into quiescence by the time that the SDSS spectrum was obtained. One way to possibly identify these quiescent systems is to identify outliers in the respective X-ray luminosities. A histogram of the X-ray luminosities for the above-mentioned object classes is shown in Figure \ref{fig:hist_Lx_contaminants}, and reveals numerous M-stars with log(L$_\text{X}$/\text{erg s$^{-1}$}) $>$ 31, as well as some WDs. While there are many M-stars with log(L$_\text{X}$/\text{erg s$^{-1}$}) $>$ 31, this might not be too surprising as flare stars have been observed to have log(L$_\text{X}$/\text{erg s$^{-1}$}) $\sim$ 32 during flaring events \citep[][]{X-ray_from_flare_stars}. What is more surprising is the fact that we find four isolated WDs which have log(L$_\text{X}$/\text{erg s$^{-1}$}) $>$ 31, as the optical spectra of these systems did not reveal any signs of accretion or binarity. However, these objects will be followed up in future studies to identify their true nature.    

\begin{table}[h]
\centering 
\caption{Observed number of contaminants per carton} 
\label{tab:contaminants}
\begin{threeparttable}
\begin{tabular}{@{}c|ccc|c}
    \hline
    \hline
    \multicolumn{1}{}{} &
    \multicolumn{3}{c}{carton}\\    
    \multicolumn{1}{}{} &
    \multicolumn{3}{c}{mwm\_erosita\_compact\_}\\
    \hline

                               & gen        &   var                  &   boss                   &   \\
Object class & (3,281)\tnote{$^\star$}   & (1,850)\tnote{$^\star$}   & (1,658)\tnote{$^\star$}  & Unique\\
    \hline
M-star                         & 1,388      & 925                    & 18                       & 1,488\\
WD                             & 7          & 1                      & 197                      & 199\\
WD + M-star\tnote{$^\dag$}     & 20         & 19                     & 3                        & 24\\
young stellar object           & 57         & 43                     & 4                        & 70\\
extragalactic\tnote{$^\ddag$}  & 115        & 103                    & 102                      & 222\\
	\hline
	\hline
\end{tabular}
\begin{tablenotes}
\item[$^\star$] total number of targets observed in carton 
\item[$^\dag$] detached binaries
\item[$^\ddag$] galaxies, AGN, and QSOs
    \end{tablenotes}
\end{threeparttable}
\end{table}   

\subsection{Uncertain cases}

There were, in addition, also several spectra that had good signal-to-noise ratio, however, a simple classification still proved rather illusive. Most of these are WDs, showing possible weak signs of accretion, and might be good period bouncer candidates (see Hernández-Díaz et al. (in prep.) for \erosita\ selected PB CV candidates observed in SDSS-V), while some might be weakly accreting mCVs. There were also several active M-stars that showed prominent and broad Balmer emission lines, however the spectra did not reveal any additional signs of accretion, such as a strong blue continuum or WD Balmer absorption. We did not include these systems in our CV list. We highlight two of these uncertain cases in Appendix \ref{appen:uncertain_cases}, Gaia DR3 3074742107676979584 and Gaia DR3 3134472183503345408, both of which are WDs, which we consider strange and merit future follow-up investigations.

\subsection{Catalogue} \label{subsec:catalogue}

A catalogue accompanying this paper was created, in which all the CVs that were identified in the three cartons that we submitted to the MWM programme, as well as those CVs identified in the BHM SPIDERS cartons, are listed, and thus consists of 587 entries. An overview of the catalogue is shown in Table \ref{tab:catalogue_columns} in the Appendix, together with a description of the columns. 
It consists of optical data from \gaia\ DR3, and X-ray data from the final processing version of the eRASS:3 data, unless otherwise stated. These eRASS:3 data will become public with \erosita\ DR2. 

Three systems have IAUNAME entries, but do not have \erosita\ detection IDs (DETUIDs). These three were amongst the systems identified in the SPIDERS cartons, and were observed in the \erosita\ Final Equatorial Depth Survey \citep[eFEDS,][]{eFEDS}. The eFEDS field observation reached a flux limit of 6.5$\times$ 10$^{-15}$ erg s$^{-1}$ cm$^{-2}$ in the 0.5-2.0 keV band, far deeper than the flux limit of eRASS:3. Indeed, we find that these three sources all have F$_\text{X}$ < 1.81 $\times$ 10$^{-14}$ erg s$^{-1}$ cm$^{-2}$, which is below the eRASS:3 flux limit. However, we include these systems as they were detected by \erosita\ and observed in SDSS-V, and we use the \erosita\ flux and position measurements from the published eFEDS catalogue. 

There are also seven systems that do not have an IAUNAME, all of which were X-ray sources that were identified in a previous reduction version of the eRASS:3 data (c020). This previous processing version of eRASS:3 was used to identify the targets that formed part of the mwm\_erosita\_compact\_boss carton, as this was the latest processing version available at the time of target selection and submission to SDSS-V. However, the later processing version, which \erosita\ DR2 is based on, does not contain these targets any longer. The likely reason for this is that all, except one, have F$_\text{X} <$ 2.0 $\times$ 10$^{-14}$ erg s$^{-1}$ cm$^{-2}$ as well as low detection likelihoods (with an average DetML $\lesssim$ 6), and subsequently were lost in the later renditions of the eRASS:3 reduction version (c030). In fact, this does seem likely to be the case, as five of these seven systems are found in eRASS:4, data which is still, at the time of writing, proprietary to the \erosita\_DE consortium.

This change in detection likelihood between processing versions should not be surprising, as it is strongly dependent on the background subtraction algorithm implemented during processing, which may be different between processing versions. These seven systems can be identified by their DETUID, which end with the sequence c020. We choose to include these systems, as they were still identified by \erosita\ and are in previous versions of the eRASS:3 data processing, with the X-ray data given for these systems coming from processing version c020. 

Regarding the distances of the systems, we decided to use the distances as determined by \citet[][]{bailer-jones_correct}, i.e. r$_\text{{geo}}$, for the remainder of our analysis, as those distances were also used in previous volume-limited studies \citep[][]{pala, inight_300pc}.
Looking at the distance distribution, we found that all of our systems have a distance < 8 kpc, with 125 being < 500 pc, 22 of which new to the literature, and seven are within 150 pc, two of which are new discoveries.    

Literature searches were done on all the systems to collect known orbital periods, as well as classify and sub-classify each system, which we list in the catalogue. From these classifications we found 14 AM CVns, four of which we identified from this work, 279 DNe, two NA and one NB novae, 30 NLs, 52 known mCVs, including one LARP that we identified, as well as 17 mCV candidates that we identified. We also found that 114 systems only had AAVSO/VSX entries, with one NA, five listed as NLs, and the rest as DNe.  

\section{Discussion and conclusions} \label{sec:discussion_conclusion}

Here we have reported on the eROSITA selected ACB candidates that were observed in SDSS-V, and which will become public in SDSS DR20. Our candidates were distributed in three cartons, two of which, mwm\_erosita\_compact\_gen and mwm\_erosita\_compact\_var, were based on Gaia DR2 positional crossmatch to the eRASS1 observations, while the third, mwm\_erosita\_compact\_boss, incorporated machine learning together with eRASS:3 and Gaia DR3 data. Amongst the three cartons were 89,337 unique targets that were identified as ACB candidates, of which the vast majority came from the mwm\_erosita\_compact\_gen carton. 

Of the 89,337 unique targets submitted, only 4,946 were observed by MJD 60708 and will form part of SDSS DR20. After visual inspection of these observations, we found 542 to be ACBs. This gives an overall success rate for the three cartons of 11.0\%. However, when looking at each carton individually, we see that the mwm\_erosita\_compact\_boss carton had by far the greatest ACB success rate, at 31.5\%, while mwm\_erosita\_compact\_var and mwm\_erosita\_compact\_gen had success rates of 10.8\% and 5.8\%, respectively (see Table \ref{tab:cv-submit_observed}). We can therefore see that candidate selection based on Bayesian statistics, which was employed for the mwm\_erosita\_compact\_boss candidates, together with the eRASS:3 improved accuracy in target positioning, seem to produce far better ACB counterpart candidates to the X-ray sources than just positional matching based on proximity and optical variability considerations. 

An additional 49 ACBs were identified by visual inspection of the SDSS-V spectra that were obtained for \erosita\ sources within three BHM SPIDERS cartons (spiders\_edr3, spiders\_ls10, and efeds), and which were not amongst the targets selected within our cartons. Thus, a total of 591 ACBs were identified as the optical counterparts to \erosita\ X-ray sources, including those from the BHM SPIDERS cartons.   

Matching our 591 sources to known XRB catalogues, we found four LMXBs when matching to \citet[][]{LMXB_cat}, while no HMXBs were found when matching to \citet[][]{HMXB_cat}. However, given that in the optical, photometrically and spectroscopically, XRBs are almost indistinguishable from CVs, it is very reasonable to expect that they could form part of our candidate selection. The X-ray information at hand also did not help distinguish these systems from CVs, as the X-ray luminosities measured for the four LMXBs were 30.9 $\lesssim$ log(L$_\text{X}$/\text{erg s$^{-1}$}) $\lesssim$ 33.0, with PSR J1023+0038 having the highest X-ray luminosity of the four. These values are well within what is expected of CVs, apart perhaps for PSR J1023+0038 which is higher than what is average for even IPs \citep[][]{schwope_first_systematic}. That being said, PSR J1023+0038 is not the most X-ray luminous object in our list, as the IP V418 Gem, had log(L$_\text{X}/\text{erg s$^{-1}$}) \sim33.15$. 

\begin{figure}
     \centering
        \includegraphics[width=\columnwidth]{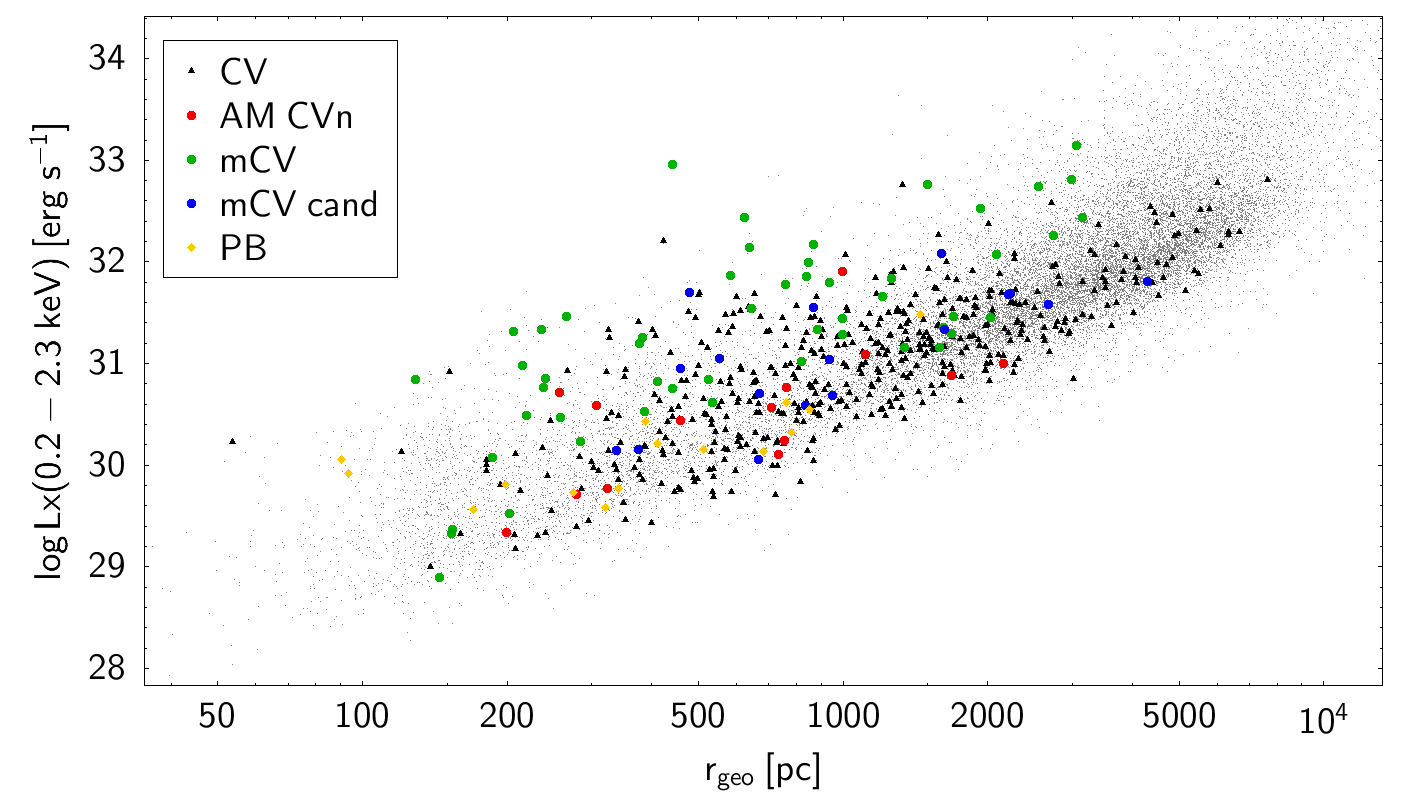}
        \caption[DR20\_Lx\_dist]{X-ray luminosity in the 0.2 -- 2.3 keV range against distance of all the CVs identified in our screening of the SDSS-V DR20 spectra of \erosita\ selected CV candidates. In all three we identify our CVs black triangles, with specific subtypes highlighted. In red we have the AM CVns, green are the known mCVs, while in blue we have the mCV candidates identified in this work, and in orange diamonds we have the period bouncers.}
        \label{fig:DR20_Lx_dist}

\end{figure}

Thus, given that four LMXBs were amongst our candidates, we therefore accept that there might be some, still unknown, XRBs that might potentially have been identified as CVs during our visual inspection, however, their inclusion will not have a significant impact on the results of this study. If there are still remaining XRBs in our CV list, they will most likely be amongst the more distant systems, as \citet[][]{LMXB_cat} and \citet[][]{HMXB_cat} both, respectively, list only two out of a combined 518 known XRBs (<0.8\%) within 500pc, while there is also an apparent overabundance of XRBs close to the Galactic Centre \citep[][]{XRBs_GC, XRBs_GC_2}. Hence, it seems unlikely, although not impossible, that undiscovered XRBs will be amongst the 22 potentially new CVs we identified within 500pc. Plotting the nominal \gaia\ distance against the X-ray luminosity, Figure \ref{fig:DR20_Lx_dist}, could prove useful to identify possible XRBs. The plot reveals $\sim$12 systems with d > 5 kpc, which might be good candidates for XRBs, although all of these systems have log(L$_\text{X}$/\text{erg s$^{-1}$}) $<$ 32.9, and thus would have to be XRBs with low X-ray luminosities. However, \citet[][]{inight_2025} lists two of these as DNe.   

Thus, after removing those four LMXBs, we are left with 587 CVs identified from the screening of the SDSS-V spectra of \erosita\ sources. Literature searches of these revealed 274 systems that were already known to the literature, and orbital periods were found for 181 of these. 

During our visual inspection, we also found 14 systems (or $\sim2.4\%$) that had an early-type CV companion, where the spectrum indicated the presence of a possible G- or K-type star. This is a surprisingly low number, as recent population studies found this fraction to be $\sim5\%$, while theory predicts this to be $\sim30\%$ \citep[][]{pala}. As CVs typically form with these early-type companions, understanding these systems is therefore crucial to understanding CV evolutionary paths, and therefore determining their orbital periods will help pin down the spectral class of the donor. However, as the companions in these systems are larger than the typical M dwarf companions, they fill their Roche-lobes at greater distances from the WD, resulting in these systems having longer orbital periods \citep[see for example][]{CV_evolution_knigge, long_CV_Porb}. 

Therefore, obtaining the orbital period using ground-based observations could prove difficult, as the orbital period might be longer than what a single observing night will allow us to monitor the system for, and hence multiple consecutive nights would be required to determine the orbital period of any such system. Such an observing stragety is costly, and will enivitably introduce uncertainties in any period found due to potential aliasing. Luckily, however, the Transiting Exoplanet Survey Satellite \citep[\tess,][]{tess} could prove invaluable in solving this problem, as it observes the same area of the sky uninterruptedly for $\sim4$ weeks at a time.  

We also identified 14 AM CVn systems, four of which are not reported in any literature searches, and thus might be new to the literature. This is a noteworthy result, as the space density of AM CVns is still uncertain by a large margin, and observation and theory disagree by an order of magnitude \citep[][]{AM_CVn_space_density, AM_CVns}. However, having an orbital period for these potentially new systems will be the surest way to definitively confirm whether they are indeed AM CVns, although their SDSS-V spectra do strongly suggest this classification. 

As the orbital periods in these systems are very short, typically <65 minutes, with HM Cancri having the shortest known orbital period of only 5.36 minutes \citep[][]{HM_cancri_discovery, HM_cancri}, using archival \tess\ observations might not be adequate, as the observing cadence might not be optimal to identify very short orbital periods. We therefore plan to obtain future ground based photometric observations in order to determine the orbital periods of these AM CVns. Confirming their nature, and thereby enlarging the sample of known AM CVns, could also help understand what their contributions could be to the background gravitational wave signal that will be observed by \lisa, as the ultra-compact AM CVns will be instrumental in the study of angular momentum loss due to gravitational radiation in close binary systems. 

One of the big outstanding questions regarding CV research, and the basis for our SDSS-V observations, is their population density, as well as their demographics, in particular, what is the ratio of magnetic to non-magnetic systems. For this we are interested in creating a 500 pc volume-limited sample. We choose to adapt the distance measurements as defined by \citet[][]{bailer-jones_correct}, as this is the same as was used in previous volume-limited CV sample studies, in particular those compiled by \citet[][]{pala} and \citet[][]{inight_300pc}, and found 124 systems with d < 500 pc. Of these 124 systems, we found 25 (or 20.0\%) that were either known, or suspected of being, mCVs. We based our classification as a likely mCV on several metrics, including the strength of \ion{He}{ii} 4686\AA, the presence of cyclotron harmonics in the spectrum, their X-ray hardness ratios, as well as their respective X-ray luminosities. This is an interesting result, as \citet[][]{pala} found that $\sim36\%$ of their 150 pc sample, consisting of 42 objects, were magnetic, while \citet[][]{inight_300pc} found that 20\% of their sample, consisting of 507 objects, were mCVs. This is in excellent agreement with our result. Relaxing the distance restriction, we found a total of 69 known, or suspected, mCVs. Identifying mCVs at greater distances are of particular interest as we try to constrain, in future studies, the contribution of mCVs to the GRXE.

However, it must be highlighted that it is possible that we could have missed some mCVs, more specifically polars when in low-states, inspecting the SDSS spectra. The reason for this is the cadence of the observations, as many systems had only one spectrum taken. A prime example of this is the study of the polar IGR J14536$-$5522 by \citet[][]{polar_low_state}. Figure 2 in their study shows two spectra of IGR J14536$-$5522, one taken during a low-state, while the other was taken during a high-state. The high-state spectrum shows very clear CV characteristics, including strong \ion{He}{ii} $\lambda$4686 emission, thus making a CV classification straightforward. However, it is immediately clear that this is not the case when the system is in a low-state, where the spectrum resembles more that of an M-star than a CV. IGR J14536$-$5522 was also observed as one of our candidates, and its spectrum would suggest that it was observed during a low-state. However, it was observed on multiple epochs, and the last spectrum did reveal something curious, a rising blue continuum with, what would otherwise be regarded as an M-star spectrum, which suggested a hot component in the system. In Figure \ref{fig:appen_145340_polar} we show the change in the flux towards shorter wavelengths in two spectra taken 40 days apart, together with the ATLAS light curve, which clearly shows the bimodal, low- and high-states, of the system. With this in mind, it is possible that some systems that we classified as M-stars, might in fact be polars observed during a low-state.

Amongst the 69 mCVs, we have 36 polars, 13 IPs, and two systems that do not have a subclassification. Excitingly, continuing with mCVs, we identified a new LARP (3eRASS J023101.9$-$585844), which lies at a distance of only 144.8 $^{+2.2}_{-1.6}$ pc, with a magnetic field of 41.3 $\pm$ 0.5 MG. With only 33 LARPs confirmed \citep[][]{PolarCat}, this discovery increases the known population by 3\%. We also identified another new CV, which is possibly a PB candidate (3eRASS J033833.9$-$332755), that lies at a distance of 138.6 $^{+3.4}_{-3.4}$ pc. \citet[][]{rodriquez_new_150pc} also found two new CVs within 150 pc using eRASS1 and \gaia\ DR3 data, and which were not in the 42 CVs found by \citet[][]{pala}. This is interesting, as both our study and that done by \citet[][]{rodriquez_new_150pc} is limited to the western Galactic hemisphere. 

Figure \ref{fig:diagnsotic_plots} gives the colour-magnitude and colour-colour diagrams, while Figure \ref{fig:DR20_Lx_dist} gives the X-ray luminosity against distance of the 587 CVs we found in this study, highlighting the AM CVns, mCVs, and mCV candidates. Looking the colour-magnitude plot, Figure \ref{fig:CM_selection}, we see that the vast majority of the systems lie between the main sequence and the WD track, with five AM CVns, two mCVs, and one mCV candidate laying very close or on the WD track. It is possible that those other CVs that lie close to the WD track could be good PB candidates, as these systems have low $\dot{\text{M}}$. A complementary paper focusing on the PB CV candidates found in \erosita\ and observed by SDSS-V will be published by Hernández-Díaz et al. (submitted), where they determine the magnetic field strength of magnetic PB candidates by analyzing the Balmer line Zeeman splitting. In their study they also report on AM CVn systems that they identified that were in their list of PB candidates.   

We also see, in both Figures \ref{fig:CM_selection} and \ref{fig:CC_selection}, that $\sim$10 systems have BP - RP $>$ 2 mag, with one of these systems being the LARP we discovered. All of these were amongst the submitted eRASS1 candidates, as a colour cut of  BP - RP $<$ 2 mag was imposed on our eRASS:3 candidate selection. In retrospect, although most CVs have BP - RP $<$ 2 mag, this decision was not the correct approach, as we are interested in identifying all CVs in the \erosita\ data, and this would automatically eliminate many low-accreting systems.     

\begin{table}[h!]   
\centering 
\caption{Success rate and predicted number of CVs in the mwm\_erosita\_compact\_boss carton based on CV likelihood} 
\label{tab:predicted_cvs_from_ML}
\begin{threeparttable}
\begin{tabular}{@{}c|ccccc}
    \hline
    \hline
    \multicolumn{1}{}{} &
    \multicolumn{5}{c}{P$_\text{cv}$} \\    
    
    \hline
            & 0.5--0.6 & 0.61--0.7 & 0.71--0.8 & 0.81--0.9 & 0.91--1   \\
    \hline        
candidates  & 6127      & 1243       & 898        & 998        & 1847       \\
observed    & 774       & 189        & 168        & 181        & 346        \\
CVs         & 32        & 43         & 46         & 77         & 325        \\
\hline
success \%  & 4.13      & 22.75      & 27.38      & 42.54      & 93.93      \\
expected    & 253       & 282        & 245        & 424        & 1734       \\
\hline
\hline
total       &           &            & 2938       &            &            \\
	\hline
	\hline
\end{tabular}
\end{threeparttable}
\end{table}

In summary, we have identified potential CVs amongst \erosita\ X-ray sources. We have identified those objects for follow-up spectroscopic observations in the framework of SDSS-V. We report on the observations obtained for the three cartons of CV candidates that we submitted to the SDSS-V MWM Survey, as well as those CVs identified in the BHM SPIDERS cartons, and found that 587 CVs. These candidates were all selected as \erosita\ X-ray sources, and are limited to the western Galactic hemisphere. This number is similar to the number of CVs found in all previous, and all-sky, SDSS observations. As the SDSS-V observations are still ongoing, Table \ref{tab:predicted_cvs_from_ML} gives an estimate of the number of CVs that could be observed by SDSS-V based on the success rate of several P$_\text{cv}$ bins for the mwm\_erosita\_compact\_boss carton, assuming all the candidates are observed. However, based on the total number of observations obtained at LCO and APO respectively, we estimate that $\sim$30.6\% of our candidates will be observed, giving a total of $\sim$900 CVs that might be observed by the end of SDSS-V observations. We plan to report on SDSS-V observations obtained after DR20 at the appropriate time.

\begin{acknowledgements}
JB acknowledges support from the Deutsche Forschungsgemeinschaft, DFG project number Schw 536/37-2. SHD acknowledges financial support from Deutsche Forschungsgemeinschaft (DFG) within research unit FOR 2990 under grant number STE\,1068/6-2. This project has received funding from the European Research Council (ERC) under the European Union’s Horizon 2020 research and innovation programme (Grant agreement No. 101020057).
GT was supported by grants IN113723, IN109723 from the Programa de Apoyo a Proyectos de Investigación e Innovación Tecnológica (PAPIIT).
Funding for the Sloan Digital Sky Survey V has been provided by the Alfred P. Sloan Foundation, the Heising-Simons Foundation, the National Science Foundation, and the Participating Institutions. SDSS acknowledges support and resources from the Center for High-Performance Computing at the University of Utah. SDSS telescopes are located at Apache Point Observatory, funded by the Astrophysical Research Consortium and operated by New Mexico State University, and at Las Campanas Observatory, operated by the Carnegie Institution for Science. The SDSS web site is \url{www.sdss.org}.

SDSS is managed by the Astrophysical Research Consortium for the Participating Institutions of the SDSS Collaboration, including the Carnegie Institution for Science, Chilean National Time Allocation Committee (CNTAC) ratified researchers, Caltech, the Gotham Participation Group, Harvard University, Heidelberg University, The Flatiron Institute, The Johns Hopkins University, L'Ecole polytechnique f\'{e}d\'{e}rale de Lausanne (EPFL), Leibniz-Institut f\"{u}r Astrophysik Potsdam (AIP), Max-Planck-Institut f\"{u}r Astronomie (MPIA Heidelberg), Max-Planck-Institut f\"{u}r Extraterrestrische Physik (MPE), Nanjing University, National Astronomical Observatories of China (NAOC), New Mexico State University, The Ohio State University, Pennsylvania State University, Smithsonian Astrophysical Observatory, Space Telescope Science Institute (STScI), the Stellar Astrophysics Participation Group, Universidad Nacional Aut\'{o}noma de M\'{e}xico, University of Arizona, University of Colorado Boulder, University of Illinois at Urbana-Champaign, University of Toronto, University of Utah, University of Virginia, Yale University, and Yunnan University.
\\
This work is based on data from eROSITA, the soft X-ray instrument aboard \SRG, a joint Russian-German science mission supported by the Russian Space Agency (Roskosmos), in the interests of the Russian Academy of Sciences represented by its Space Research Institute (IKI), and the Deutsches Zentrum für Luft- und Raumfahrt (DLR). The \SRG spacecraft was built by Lavochkin Association (NPOL) and its subcontractors, and is operated by NPOL with support from the Max-Planck Institute for Extraterrestrial Physics (MPE).

The development and construction of the eROSITA X-ray instrument was led by MPE, with contributions from the Dr.\ Karl Remeis Observatory Bamberg, the University of Hamburg Observatory, the Leibniz Institute for Astrophysics Potsdam (AIP), and the Institute for Astronomy and Astrophysics of the University of T\"ubingen, with the support of DLR and the Max Planck Society. The Argelander Institute for Astronomy of the University of Bonn and the Ludwig Maximilians Universit\"at Munich also participated in the science preparation for eROSITA.
\end{acknowledgements}

\section*{Data availability}\label{sec:data_avail}
The catalogue is available at the CDS via anonymous ftp to cdsarc.cds.unistra.fr (130.79.128.5) or via https://cdsarc.cds.unistra.fr/viz-bin/cat/J/A+A/vol/page. SDSS-V DR20 and \erosita\ DR2 data will be publicly available at the end of the proprietary period, while the other data used in this study are available from the sources referenced in the text.

\bibliographystyle{aa}
\bibliography{bibliography}   

\begin{appendix}

\onecolumn
\section{Diagnostic plots}

\begin{figure}[ht!] 
  \centering
\begin{subfigure}[]{0.9\textwidth}
    \centering
\includegraphics[width=0.9\textwidth]{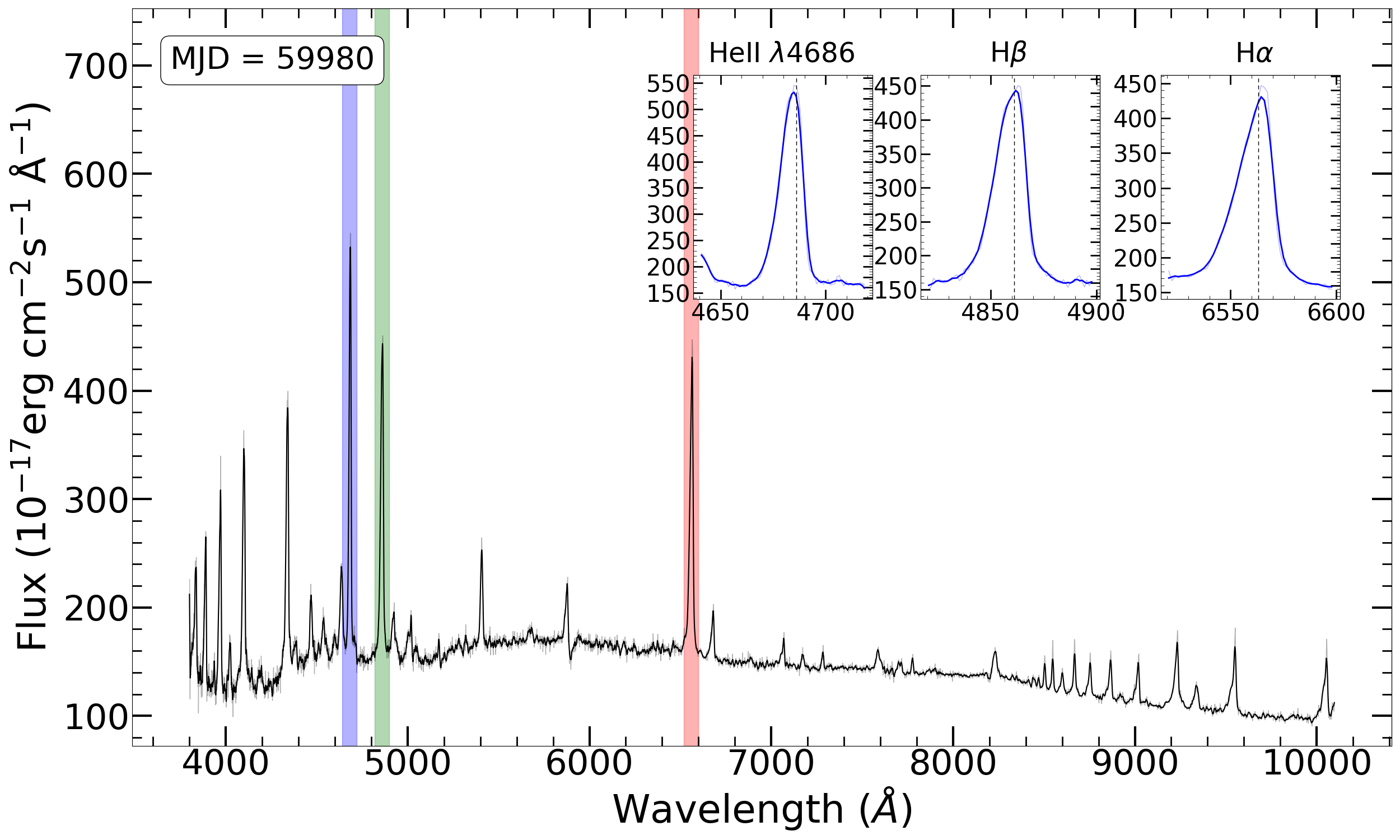}
    \caption{SDSS-V spectrum}\label{fig:V1309_Ori_SDSS_V_spectrum}
\end{subfigure}%

\vspace{1ex}

\begin{subfigure}[]{0.33\textwidth}
    \centering
\includegraphics[width=0.9\linewidth]{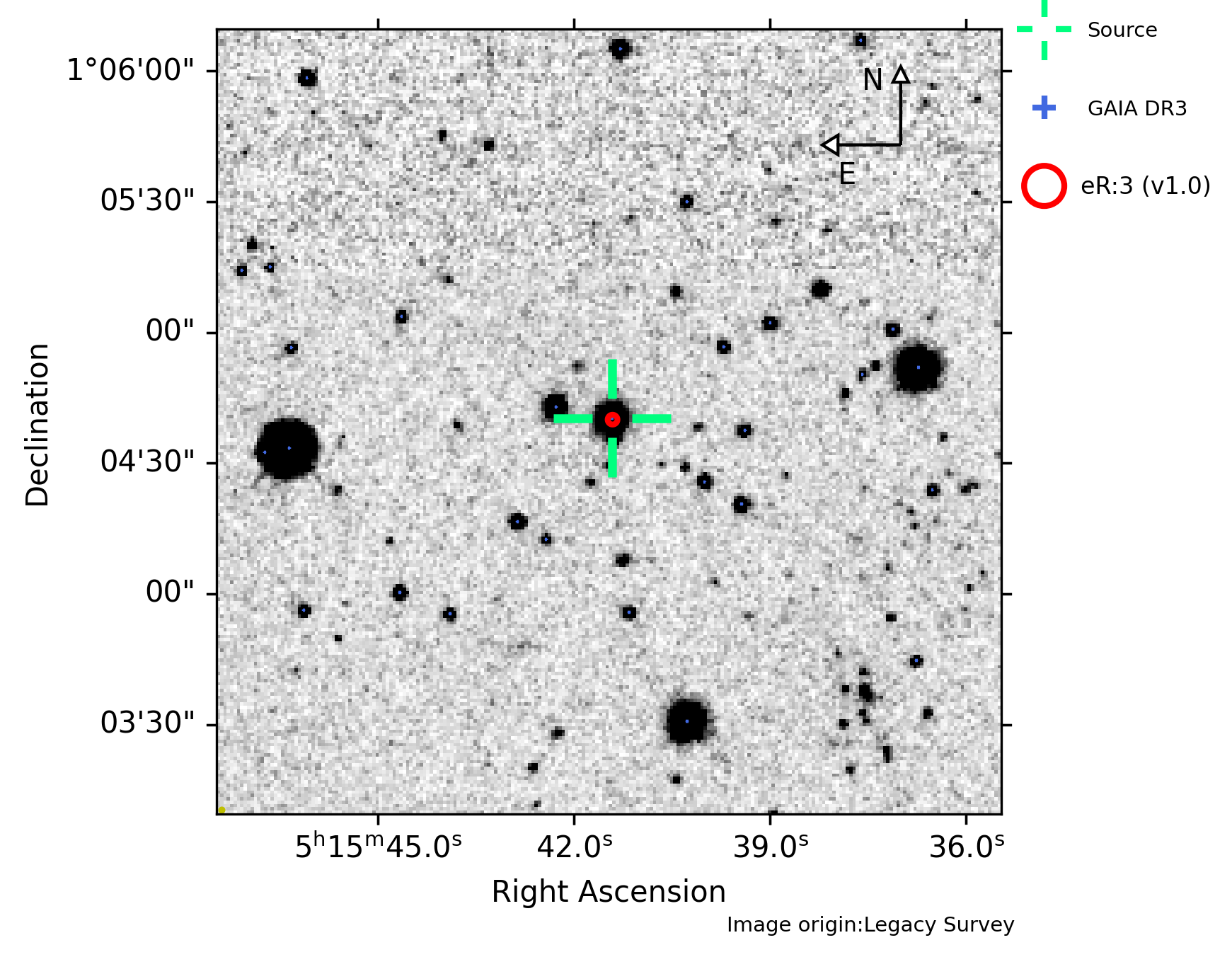}
   \caption{Finding chart}\label{fig:V1309_Ori_finding_chart}
   \end{subfigure}  
 \begin{subfigure}[]{0.33\textwidth}
\includegraphics[width=0.9\linewidth]{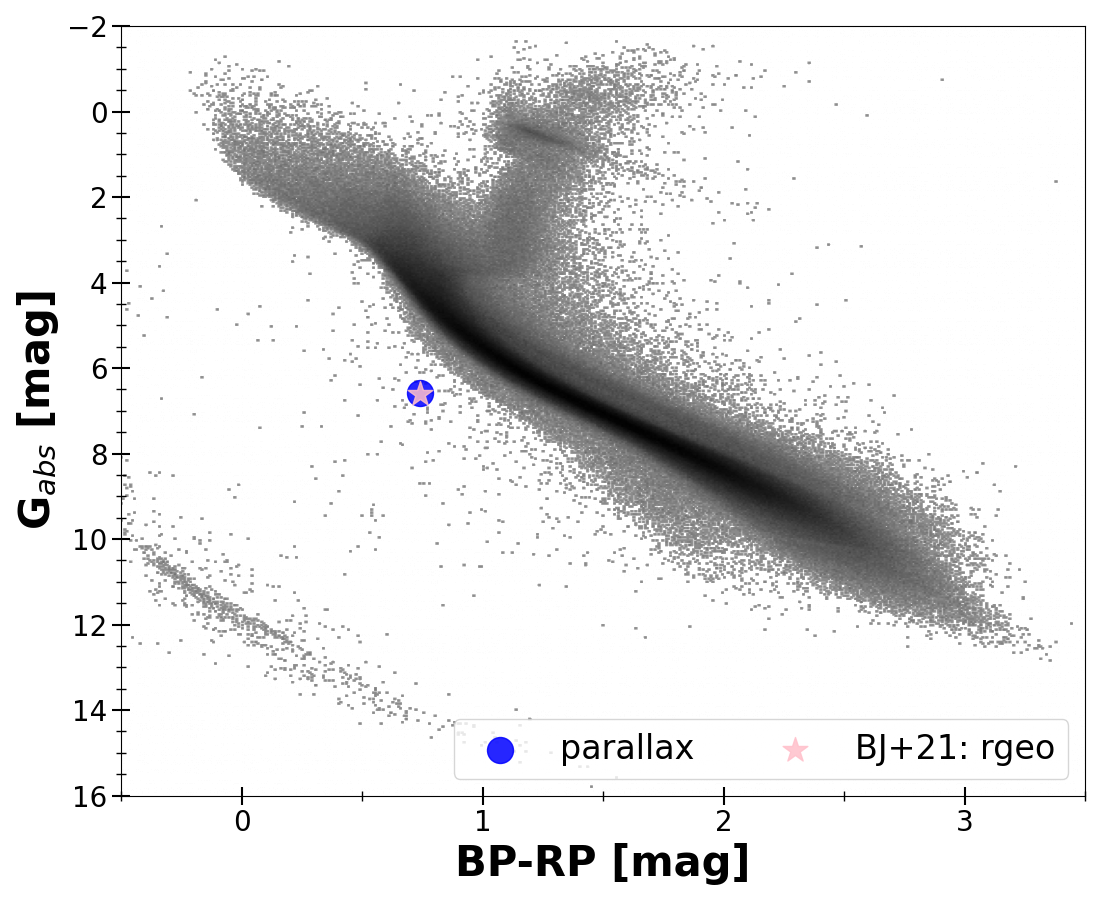}
  \caption{Colour-magnitude diagram}\label{fig:V1309_Ori_colour_mag}
  \end{subfigure}
  \begin{subfigure}[]{0.33\textwidth} 
\includegraphics[width=0.9\linewidth]{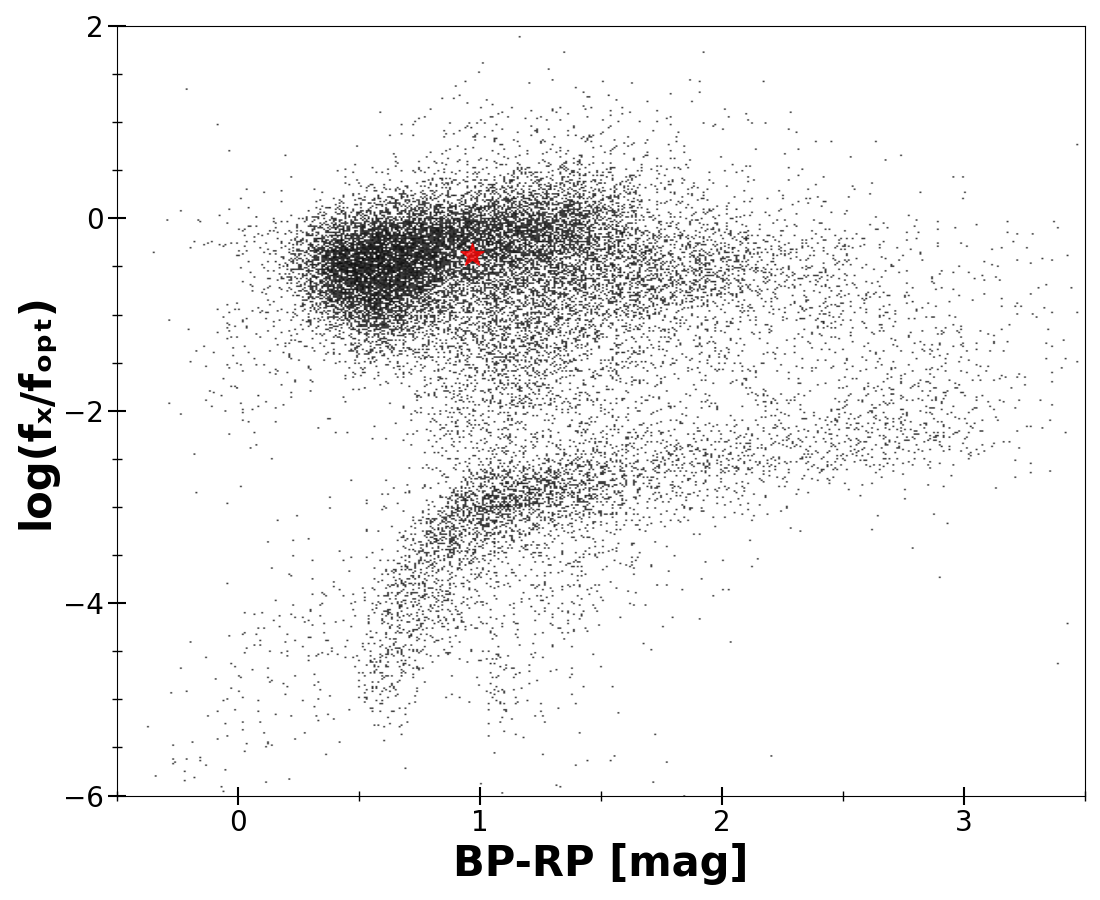}
    \caption{Colour-colour diagram}\label{fig:V1309_Ori_colour_colour}
  \end{subfigure}  

\vspace{1ex}

\begin{subfigure}[]{0.33\textwidth}
    \centering
\includegraphics[width=0.9\linewidth]{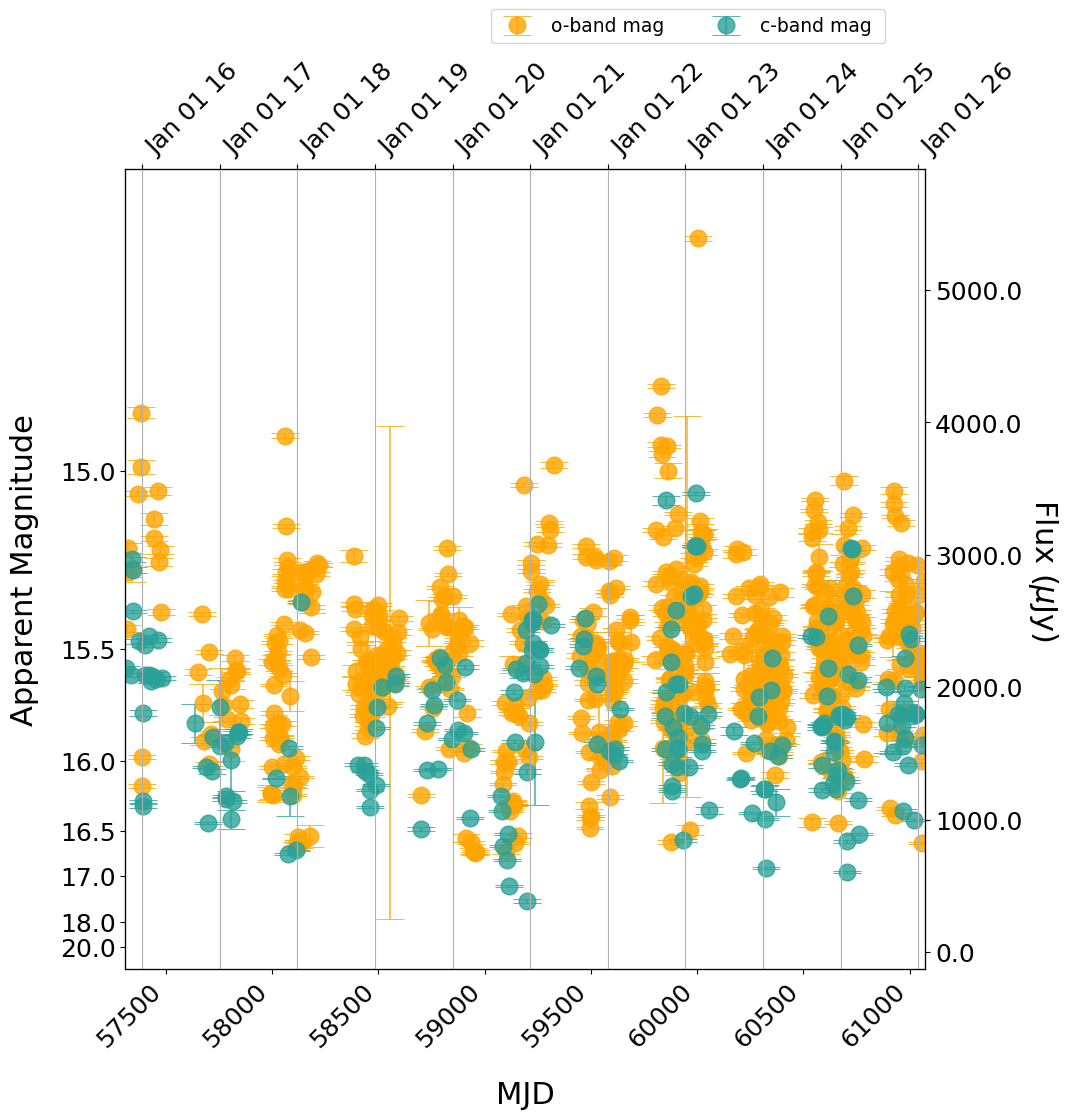}
   \caption{ATLAS light curve}\label{fig:V1309_Ori_ATLAS_LC}
   \end{subfigure}  
 \begin{subfigure}[]{0.55\textwidth}
\includegraphics[width=0.9\linewidth]{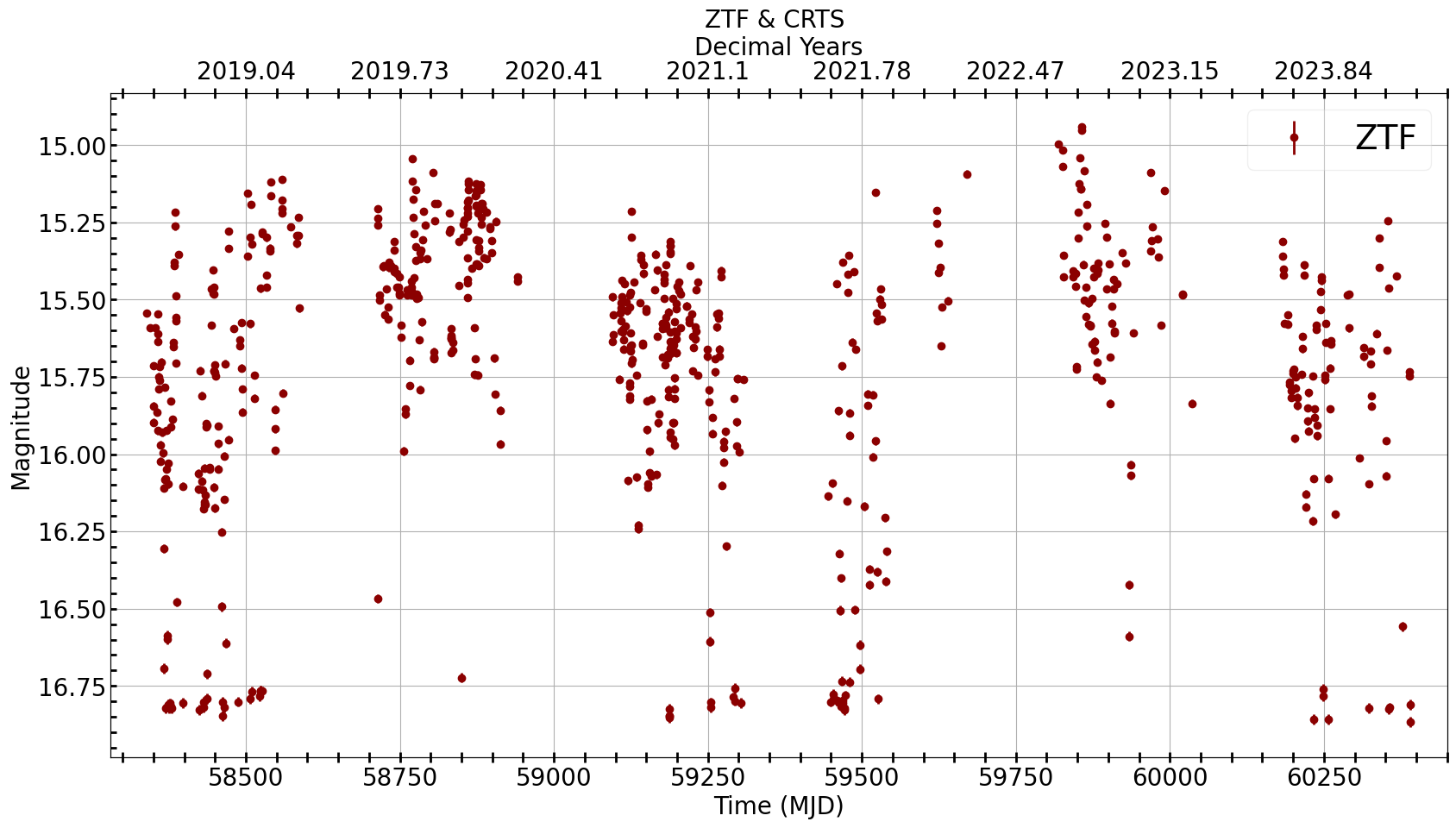}
  \caption{CRTS/ZTF light curve}\label{fig:V1309_Ori_ZTF}
  \end{subfigure}

\caption{Diagnostic plots used to determine the nature of a source, here for the known polar V1309 Ori (3eRASS J051541.4+010439), as an example. The size of the red circle in the finding chart is related to the uncertainty in the eRASS:3 X-ray positioning.}\label{fig:V1309_Ori_diagnostic}
\end{figure}

\twocolumn

\section{Catalogue}\label{appen:catalogue}

   \begin{table*}[]
    \centering 
    \caption{Description of the columns in the catalogue of CVs identified from \erosita\ selected candidates observed in during SDSS DR20} 
    \label{tab:catalogue_columns}
    \begin{threeparttable}
    \begin{tabular}{ccl}
    \hline
    \hline
    Column name             & Unit   & Description \\  
    \hline
     IAUNAME     &     & Name of system \\
     SDSS\_ID    &       & ID in SDSS \\
     DETUID &   &   ID in \erosita\ catalogues \\
     \gaia\_DR3\_ID  &   &   ID in \gaia\ DR3 \\
     RA\_eRO  & degrees   & Right ascension in eRASS:3 \\
     DEC\_eRO  & degrees   & Declination in eRASS:3 \\
     RA\_\gaia\_DR3  & degrees   & Right ascension in \gaia\ DR3 \\
     DEC\_\gaia\ DR3  & degrees   & Declination in \gaia\ DR3 \\
     RA\_ICRS     & degrees & Right ascension in the International Celestial Reference System \\
     DE\_ICRS     & degrees & Declination in the International Celestial Reference System \\
     eRO\_FLUX  &  erg s$^{-1}$ cm$^{-2}$   & X-ray flux measured in eRASS:3 in the 0.2-2.3 keV band \\
     eRO\_FLUX\_ERR  &  erg s$^{-1}$ cm$^{-2}$   & Error in the X-ray flux measured in eRASS:3 in the 0.2-2.3 keV band \\
     G$_\text{mag}$  & mag  &   Mean \gaia\ G band magnitude \\
     G$_\text{mag}$\_ERR  & mag  &  Error in the mean \gaia\ G band magnitude\\
     BP-RP  & mag  &   Difference between \gaia\ B-band and \gaia\ R-band magnitudes \\
     BP-RP\_ERR  & mag  &   Error in the BP-RP magnitudes \\
     Distance   & pc    &   r$_{\text{geo}}$ distance based on \citet[][]{bailer-jones_correct}  \\
     Distance\_LOWER   & pc    & Lower limit of r$_{\text{geo}}$ distance measurement  \\
     Distance\_UPPER   & pc    & Upper limit of r$_{\text{geo}}$ distance measurement  \\
     G$_{\text{abs}}$ & mag & Absolute G-magnitude based on G$_\text{mag}$ and r$_{\text{geo}}$ distance \\
     logL$_{\text{X}}$  & erg s$^{-1}$   & Logarithm of the X-ray luminosity based on eRO FLUX and r$_{\text{geo}}$ distance\\
     log(F$_{\text{X}}$/F$_\text{{opt}}$)   &   & Logarithm of F$_{\text{X}}$ to optical using G$_\text{mean}$\tnote{1}\\ 
     HR\_P12 &   & Hardness ratio between the eRASS:3 0.2–0.5 keV and 0.5–1.0 keV bands\tnote{2}\\
     HR\_P12\_ERR&   & Error in HR P12\\
     HR\_P23 &   & Hardness ratio between the eRASS:3 0.5–1.0 keV and 1.0–2.0 keV bands\\
     HR\_P23\_ERR&   & Error in HR P23\\
     P$_\text{orb}$ & hours & Orbital period of system \\
     CV\_TYPE    & & Main type that the CV belongs to \\
     CV\_SUBTYPE & & Subclass of CV \\ 
     REFERENCES & & List of references used to find orbital period and/or CV subtype \\
	\hline
	\hline
    \end{tabular}
    \begin{tablenotes}  
    \item[1] log(F$_{\text{X}}$/F$_\text{{opt}}$) = log$_{10}$F$_\text{X}$ + (G$_\text{mag}$/2.5) + 4.86
    \item[2] HR = (H-S)/(H+S), with H being the hard, and S being the soft bands, respectively
    \end{tablenotes}
    \end{threeparttable}
    \end{table*} 
    
For each entry we list various unique identifiers, which include the IAUNAME, SDSS\_ID, eROSITA\_DETUID, as well as the \gaia\_DR3\_ID. We also give the eRASS:3 coordinates (RA\_eRO, DEC\_eRO), \gaia\ DR3 coordiantes (RA\_\gaia\_DR3, DEC\_\gaia\_DR3), as well as the ICRS coordinates (RA\_ICRS RA, DE\_ICRS). 

Three systems have IAUNAME entries, but do not have DETUIDs. These three were amongst the systems identified in the SPIDERS cartons, and were observed in the \erosita\ Final Equatorial Depth Survey \citep[eFEDS,][]{eFEDS}. The eFEDS field observation reached a flux limit of 6.5$\times$ 10$^{-15}$ erg s$^{-1}$ cm$^{-2}$ in the 0.5-2.0 keV band, far deeper than the flux limit of eRASS:3. Indeed, we find that these three sources all have F$_\text{X}$ < 1.81 $\times$ 10$^{-14}$ erg s$^{-1}$ cm$^{-2}$, which is below the eRASS:3 flux limit. However, we include these systems as they were detected by \erosita\ and observed in SDSS-V, and use the \erosita\ flux and position measurements from the published eFEDS catalogue. 

There are also seven systems that do not have an IAUNAME, all of which were X-ray sources that were identified in a previous reduction version of the eRASS:3 data (c020). This previous processing version of eRASS:3 was used to identify the targets that formed part of the mwm\_erosita\_compact\_boss carton, as this was the latest processing version available at the time of target selection and submission to SDSS-V. However, a later processing version, and that which \erosita\ DR2 is based on, does not contain these targets any longer. The likely reason for this is that all, except one, have F$_\text{X} <$ 2.0 $\times$ 10$^{-14}$ erg s$^{-1}$ cm$^{-2}$ as well as low detection likelihoods (with an average DetML $\lesssim$ 6), and subsequently were lost in the later renditions of the eRASS:3 reduction version (c030). In fact, this does seem likely to be the case, as five of these seven systems are found in eRASS:4, data which is still, at the time of writing, proprietary to the \erosita\_DE consortium.

The columns labelled eRO\_FLUX and eRO\_FLUX\_ERR give the eRASS:3 flux measurement in the 0.2–2.3 keV band, and its error, respectively. The mean \gaia\ G-band magnitude is given in column G$_\text{mag}$, with the uncertainty given in G$_\text{mag}$\_ERR, while BP-RP is the difference between the \gaia\ B-band and R-band magnitudes, with its error given in BP-RP\_ERR. 

For the distances of the systems, we find some discrepancies between the measurements determined using the \gaia\ inverse parallax distances, and the r$_\text{{geo}}$ distances \citep[][]{gaia_dist_bailer_jones}. We found some systems that had negative parallax measurements, which does not make sense, and hence caused doubt for the inverse parallax results. We therefore decided to use the distances as determined by \citet[][]{gaia_dist_bailer_jones}, i.e. r$_\text{{geo}}$. Therefore, three distances are given, all derived from \citet[][]{gaia_dist_bailer_jones}, with the DISTANCE column being the nominal distance (r$_\text{geo}$), while DISTANCE\_LOWER and DISTANCE\_UPPER being the lower and upper limits of the distance, respectively. 

The absolute G-band magnitude is given in the G$_\text{abs}$ column, which was calculated using G$_\text{mag}$ and the nominal \gaia\ distance, i.e. DISTANCE. The logarithm of the X-ray luminosity in the 0.2-2.3 keV band of the system is given in the log(L$_\text{X}$) column, and is calculated using DISTANCE and eRO\_FLUX. No correction for interstellar absorption was made in calculating G$_\text{abs}$ or log(L$_\text{X}$). We also calculated the X-ray to optical flux ratio, with the logarithm of this value given in log(F$_\text{X}$/F$_\text{opt}$) column, and is defined as log(F$_{\text{X}}$/F$_\text{{opt}}$) = log$_{10}$(eRO\_FLUX) + (G$_\text{mag}$/2.5) + 4.86. 

We also list two hardness ratios (HR) together with their respective errors, with HR = (H-S)/(H+S), with H being the hard, and S being the soft bands, respectively. The first of these is HR\_P12, with S = 0.2–0.5 keV and H = 0.5–1.0 keV bands, respectively, with HR\_P12\_ERR being the uncertainty. Next is HR\_P23 with S = 0.5–1.0 keV and H = 1.0–2.0 keV bands, respectively, and its associated uncertainty, HR\_P23\_ERR. 

The PERIOD column consists of the orbital periods we found from literature searches, given in hours. From the literature search we were also able to obtain information on the main CV type and CV subtype classifications of many systems, with Table \ref{tab:cv_classification} giving a summary of the classifications. The main classification is given in the CV\_TYPE column, with the main classifications being DN = dwarf nova, NL = novalike, NA = fast nova, NB = slow nova, mCV = magnetic CV, as well as AM CVn systems. An additional classification, mCV\_cand, is reserved for those systems that were identified in our visual inspection of the SDSS-V spectra, light curves as well as optical and X-ray data, in particular the HRs, as likely mCVs, or mCV candidates. However, these need to be confirmed by follow-up observations. 

In the CV\_SUBTYPE column we list the subclassification of the system, which for the DNe we have subtypes UG = U Gem type, SU = SU UMa, SS = supersoft X-ray source, WZ = WZ Sge type system and UGZ = Z Cam system. We also have PB = period bouncer, which are either known or candidate period bouncer CVs, and were mainly identified amongst the DN, however a few are also amongst the NL systems. For NLs these subclassification types are VY = VY Scl systems, UX = UX UMa type systems, and SW = SW Sex systems. Lastly, for the mCVs, we have polar, LARP and IP. 

It will also be noticed that in many cases the subclass is not unique, as in many cases different subclasses are given by different references. We therefore list all the subclasses found per target from the literature review, with the first entry being the most frequently given. However, here we also list possible alternative main classification, as some references claim evidence to support these alternative classifications. This was mostly related to the PB and IP entries found in the CV\_SUBTYPE column, although some mixed classification also occurred between the NLs and DNe. 

Lastly, in the REFERENCES column we list the references used to obtain the aforementioned periods and classifications. These references are listed in a separate file that accompanies the catalogue. It must be noted that six systems are listed with a reference, but without a period or CV classification entry, as these systems were listed in the respective references as a CV, however contained neither a period nor a classification.

\begin{table*}[]
\centering 
\caption{CV main and subtype classification} 
\label{tab:cv_classification}
\begin{threeparttable}
\begin{tabular}{@{}c|c|c}
    \hline
    \hline
CV\_TYPE   & CV\_SUBTYPE   &   Description  \\
\hline
          &   UG (U Gem)  & classical dwarf nova    \\
          &   SU (SU UMa) & dwarf nova with superoutbursts\\
DN        &   SS          & supersoft X-ray source\\
          &   WZ (WZ Sge) & low accretion-rate SU UMa\\
          &   UGZ (Z Cam) & dwarf nova with standstills\\
\hline
          &   PB          & period bouncer\\
\hline
          &   VY (VY Scl) & novalike with occasional deep low states\\
NL        &   SW (SW Sex) & high mass-transfer rate novalike with peculiar emission lines\\
          &   UX (UX UMa) & stable high-state novalike\\
\hline
Nova      &   NA          & fast nova \\
          &   NB          & slow nova \\
\hline
          &   IP          & intermediate polar\\
mCV       &   Polar       & highly magnetic, synchronized binary\\
          &   LARP        & low accretion rate polar\\
\hline
mCV\_cand &               & mCV candidate identified in this work \\ 
          &               & \\
\hline  
AM\_CVn   &               & ultracompact He transferring binary\\
	\hline
	\hline  
\end{tabular}
\end{threeparttable}
\end{table*} 

\clearpage

\section{Results}

\begin{figure}[h!]
        \centering
        \includegraphics[width = \columnwidth]{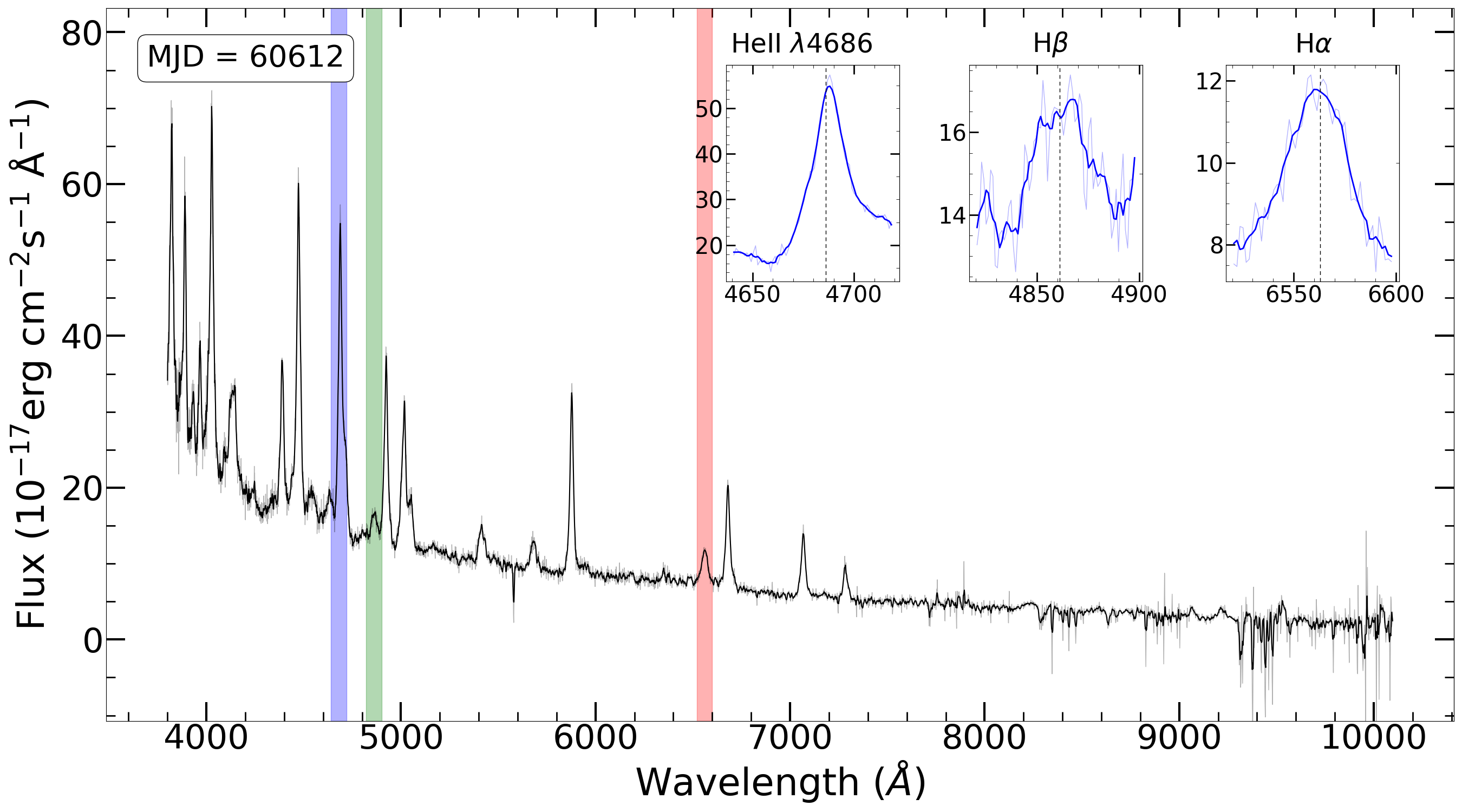}
        
        \caption[AM CVn J080449]{Spectrum of the AM CVn 3eRASS J080449.4$+$161625, with strong \ion{He}{i} and \ion{He}{ii} emission lines; however, weak Balmer emission lines are present as well. }
        \label{fig:appen_080449_Am_CVn}
\end{figure}

\begin{figure}[h!]
        \centering
        \includegraphics[width = \columnwidth]{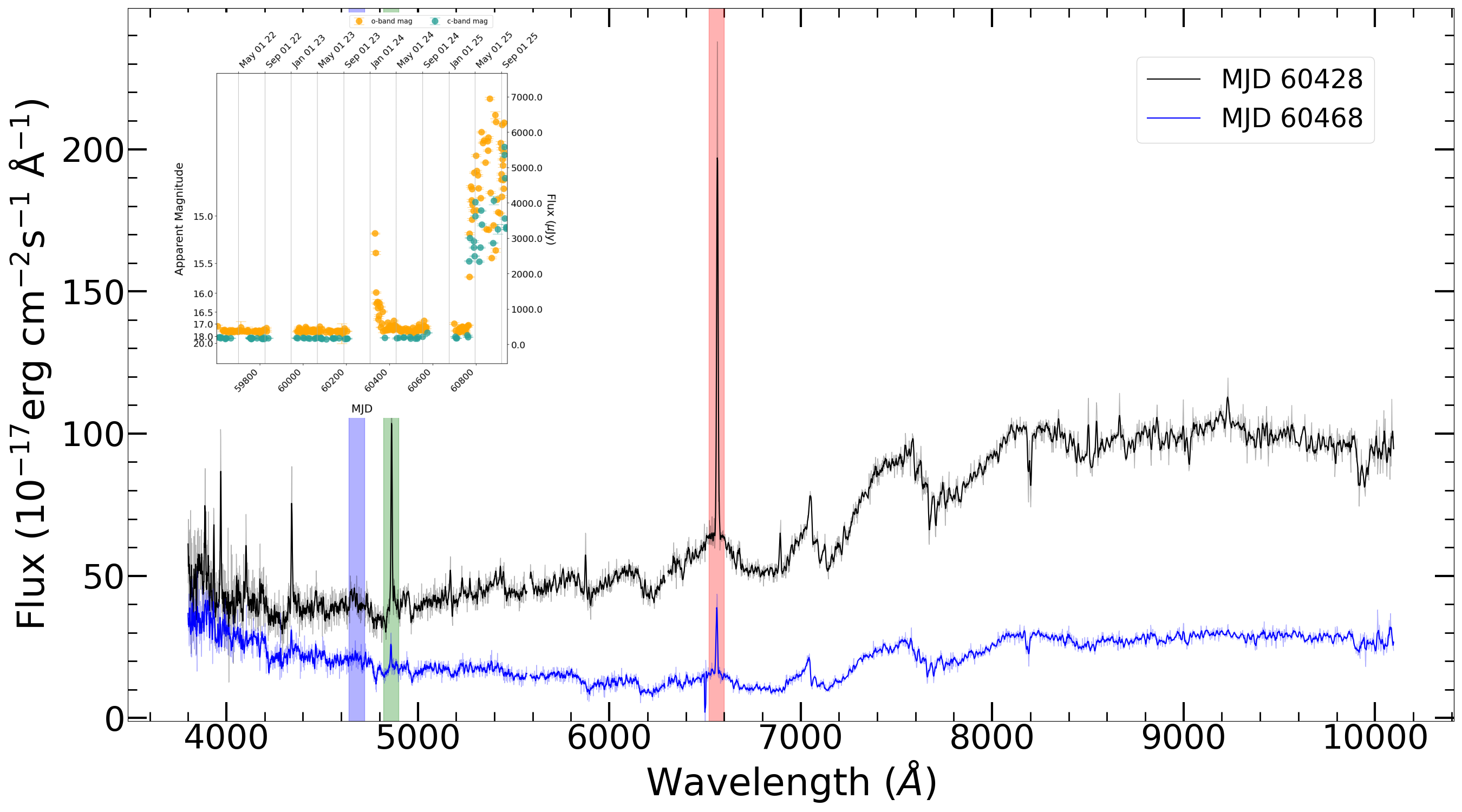}
        
        \caption[polar J145340]{Two spectra, separated by 40 days, of the polar 3eRASS J145340.8$-$552139 (IGR J14536$-$5522). Note the increase in flux towards shorter wavelengths in the MJD 60468 spectrum. The inset plot shows the ATLAS light curve, where the bimodal nature of the system is clearly seen.}
        \label{fig:appen_145340_polar}
\end{figure}

\begin{table}[h!]
\centering 
\caption{CVs with early type donors} 
\label{tab:early_type_donors}
\begin{threeparttable}
\begin{tabular}{@{}c||c}
    \hline
    \hline
    \multicolumn{1}{c}{IAU name} &  
    \multicolumn{1}{c}{IAU name} \\
    \hline
3eRASS J040435.4-310347 & 3eRASS J072053.9+341246  \\
3eRASS J044348.2-594908 & 3eRASS J072615.3-102337  \\
3eRASS J050906.7-032507 & 3eRASS J082325.0-002056  \\
3eRASS J060501.5-243747 & 3eRASS J090757.4-323859  \\
3eRASS J061916.2-493906 & 3eRASS J091410.8+013730  \\
3eRASS J062556.4-080223 & 3eRASS J134504.5+004253  \\
3eRASS J063026.9-652950 & 3eRASS J214216.5-674759  \\
	\hline
	\hline
\end{tabular}

\end{threeparttable}
\end{table}  

\begin{figure}[h!]
        \centering
        \includegraphics[width = \columnwidth]{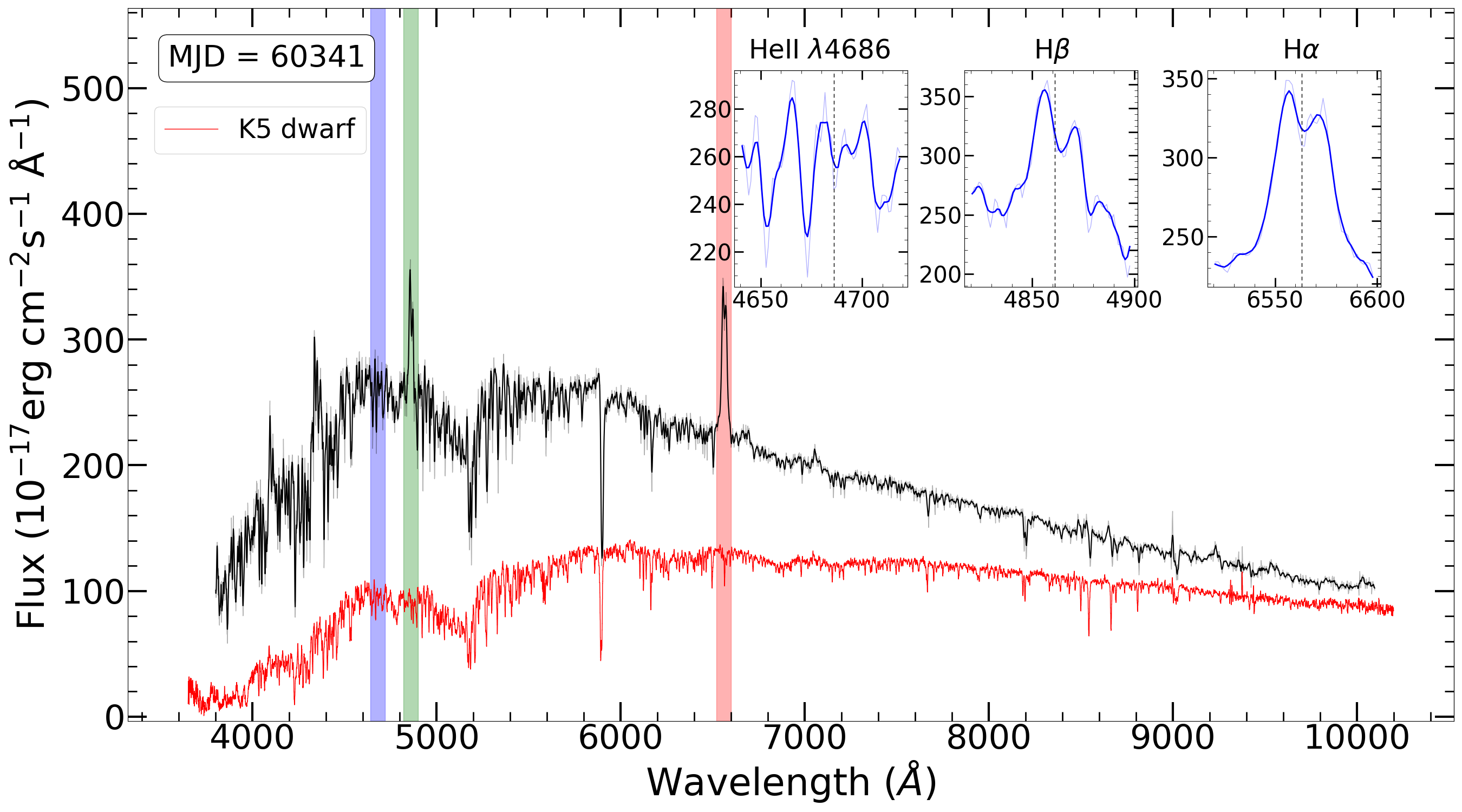}
        \caption[large secondary]{Spectrum of 3eRASS J063026.9-652950, in black, while in red we show a template spectrum of a K5 dwarf star for illustrative purposes. Note the very strong Mg absorption complex at $\lambda\sim$5170\AA, as well as the Na D absorption at $\lambda\sim5895$\AA. The inset plots focus on the respective areas in the spectrum of J063027.19$-$652950.23. The flux of the K5 star was shifted by an arbitrary amount to have a similar scale as that of the target spectrum.}
        \label{fig:appen_063027.19_G_K_companion}
\end{figure}

\clearpage

\section{New CVs within 15 0pc} \label{appen:Cvs_150pc}
\subsubsection{3eRASS J023101.9$-$585844} \label{subsubsec:LARP}

\begin{figure*}
     \centering
     \begin{subfigure}[b]{0.64\textwidth}
         \centering
         \includegraphics[width=\textwidth]{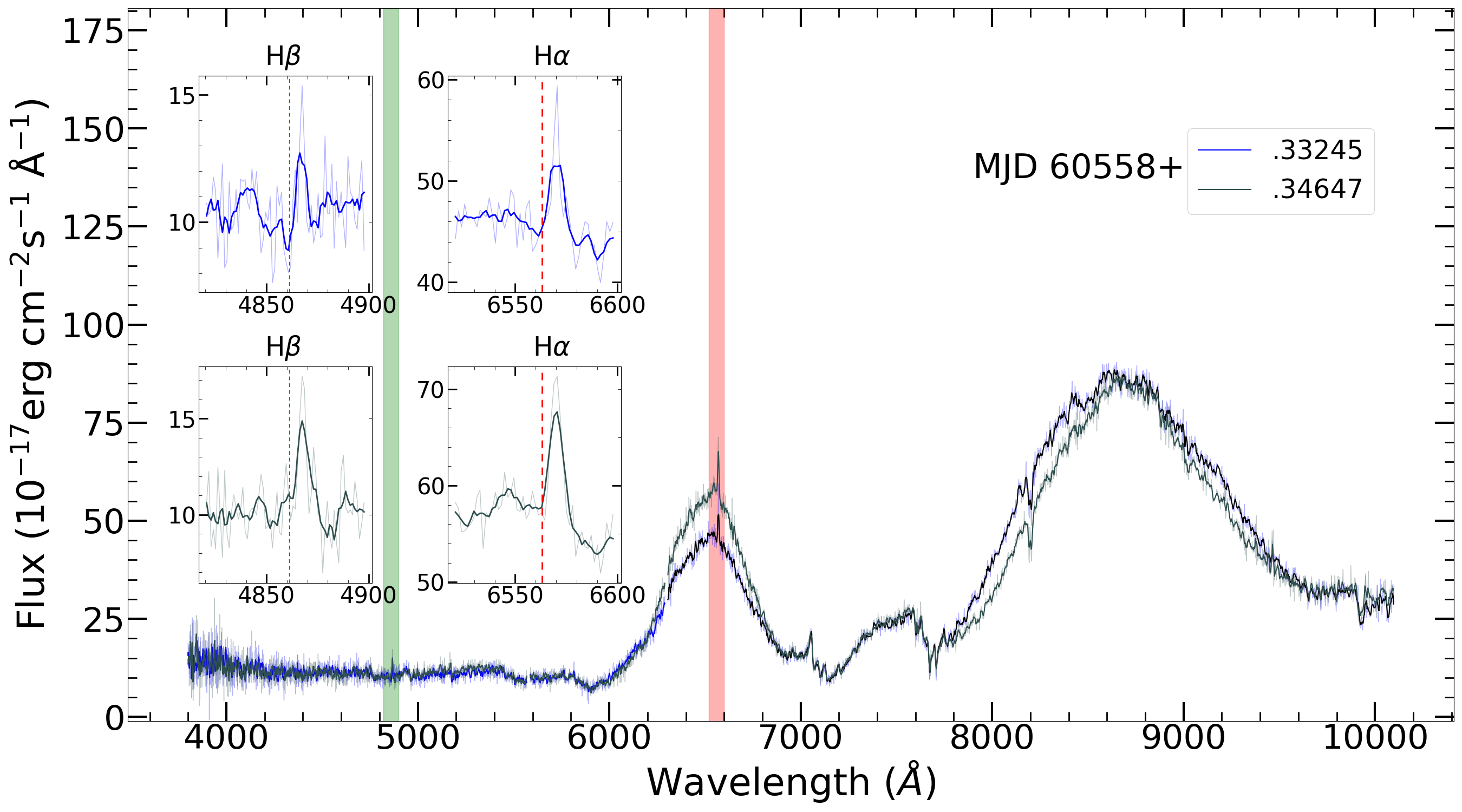}
         \caption{SDSS-V spectrum}
         \label{fig:spec_J0231}
     \end{subfigure}
     \hfill
     \begin{subfigure}[b]{0.35\textwidth}
         \centering
         \includegraphics[width=\textwidth]{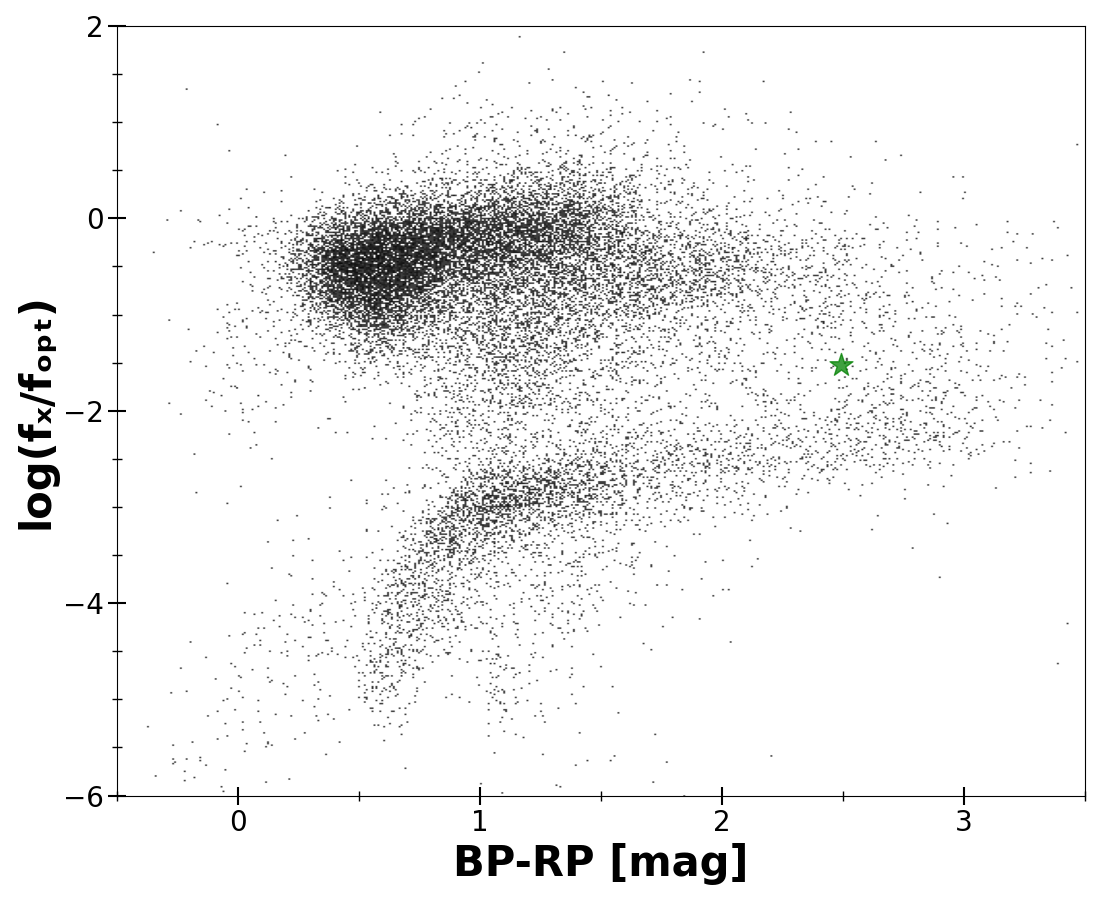}
         \caption{Colour-colour plot}
         \label{fig:CC_J0231}
     \end{subfigure}
        \caption{(a) SDSS-V sub-spectra, obtained on MJD 60558 and taken 900s apart, of 3eRASS J023101.9$-$585844 showing two prominent cyclotron harmonics, together with weak Balmer H$\alpha$ and H$\beta$ emission, which are highlighted in the inset plots. Note the change in the cyclotron harmonics between the two exposures, with the first exposure in blue and the second exposure shown in gray. No vertical offset between the two spectra was added. (b) Position of the system in the colour-colour diagram.}
        \label{fig:J0231}
\end{figure*}

The first system in question is 3eRASS J023101.9$-$585844, which lies at a distance of 144.8 pc. Looking at the optical spectrum, Figure \ref{fig:spec_J0231}, we immediately see two strong cyclotron harmonics between $\sim$6000\AA\ - 7000\AA, and $\sim$7800\AA\ - 9600\AA, respectively, indicating that the WD in the system has a strong magnetic field, with no, or very weak, signs of accretion. This is typical of low accretion rate polars (LARPs). 
These systems differ from ordinary CVs as the companion star is, in at least some of these systems, likely Roche-lobe underfilling, and instead we have wind accretion from the companion onto the WD \citep[][]{LARPS_underfil}, thus resulting in these systems sometimes being referred to as pre-polars \citep[][]{PREPs}. The X-ray emission in these systems is typically due to the coronal activity of the M-star, while in polars hard X-ray emission originates from thermal Bremsstrahlung radiation at the accretion shock front near the surface of the WD, and soft X-rays are emitted from the WD photosphere as reprocessed hard X-rays.  

Looking again at Figure \ref{fig:spec_J0231}, plotted are two sub-spectra taken 900s apart, which clearly show changes in the intensity of the two harmonics thus suggesting variability as a function of the orbital phase. The spectrum does, however, also show slightly redshifted H$\alpha$ and H$\beta$ emission, together with late-type stellar features from the companion such as the \ion{K}{i} doublet absorption at $\lambda\lambda$ $\sim$7665\AA\ and $\sim$7699\AA, as well as TiO and VO molecular bands. Looking at the X-ray properties of the system, we find that it has log(L$_\text{X}$) = 28.96 erg s$^{-1}$, and hardness ratios HR\_P12 = 0.79 $\pm$ 0.19, and HR\_P23 = -0.39 $\pm$ 0.25.  We also see, in Figure \ref{fig:CC_J0231}, that the system appears very red, with BP-RP = 2.59 mag, and lies slightly above the stellar branch in the colour-colour diagram, having log(F$_\text{X}$/F$_{\text{opt}}$) = -1.57.

To determine the magnetic field strength of the WD we first removed the contribution of the M-star from the spectrum. This was done by fitting a scaled template of an M5 star and subtracting it from the spectrum, hence leaving the emission due to cyclotron beaming. We then compared this to isothermal cyclotron models \citep[][]{isothermal_cyclotron}, and found a magnetic field strength of B = 41.3 $\pm$ 0.5 MG, with the uncertainty being due to the unknown viewing angle, i.e. binary orbital phase, at which the spectrum was obtained.

\subsubsection{3eRASS J033833.9$-$332755} \label{subsubsec:new_PB}

\begin{figure*}
     \centering
     \begin{subfigure}[b]{0.64\textwidth}
         \centering
         \includegraphics[width=\textwidth]{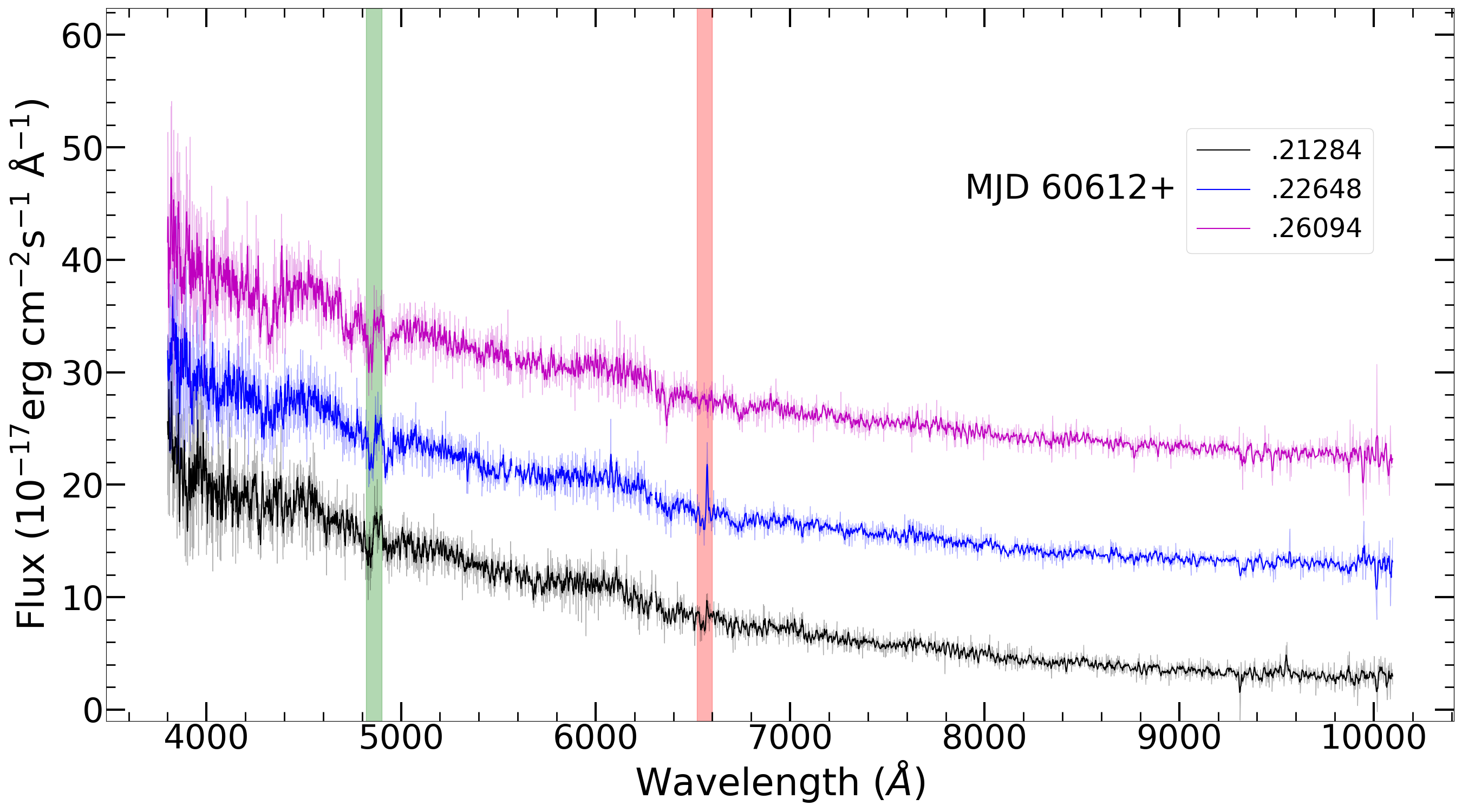}
         \caption{SDSS-V spectrum}
         \label{fig:spec_J03384}
     \end{subfigure}
     \hfill
     \begin{subfigure}[b]{0.35\textwidth}
         \centering
         \includegraphics[width=\textwidth]{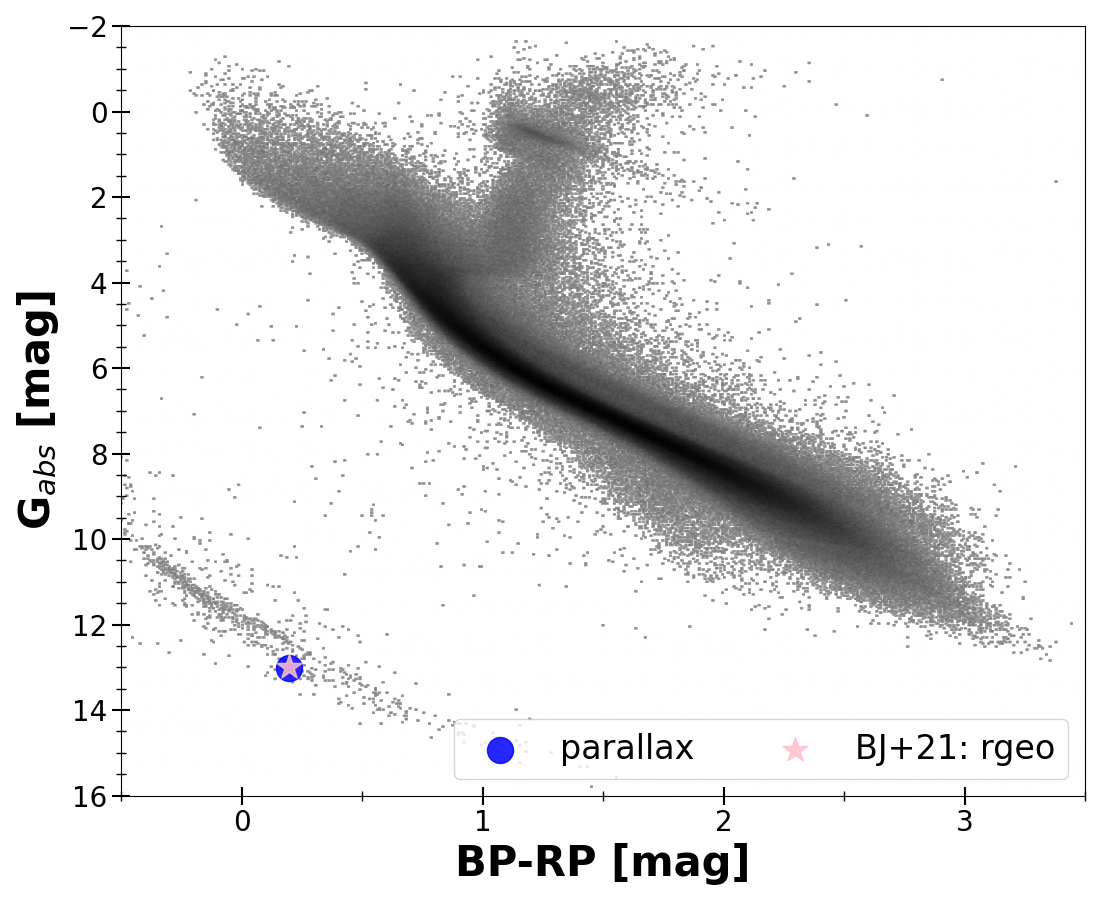}
         \caption{Colour-magnitude plot}
         \label{fig:CM_J03384}
     \end{subfigure}
        \caption{(a) Three SDSS-V sub-spectra, obtained on MJD 60612 and taken 900 s apart, of 3eRASS J033833.9$-$332755. The regions surrounding H$\alpha$ and H$\beta$ are highlighted by the red and green shaded areas, respectively. Each spectrum reveals a rising blue continuum, with strong Balmer H$\beta$ and H$\gamma$ absorption lines, indicative of a WD. However, the H$\beta$ absorption line has a very prominent and broad emission core. Transient H$\alpha$ emission is also seen. The spectra also reveal possible cyclotron emission between $\lambda\lambda$ $\sim$ 5700\AA - 6200\AA. To increase readability, a vertical offset of 10 and 20 units were applied to the observations obtained on MJD 60612.22648 and MJD 60612.26094, respectively. (b) Position of the system in the colour-magnitude diagram places the system on the WD track. For further details, see Hernández-Díaz et al. (submitted).}
        \label{fig:J03384}
\end{figure*}

The second new system that we found is 3eRASS J033833.9$-$332755, which lies at a distance of 138.6 pc, and was also identified as a PB candidate by Hernández-Díaz et al. (submitted). The system was observed on MJD 60612, with three sub-exposures obtained during this observation, each with a 900s exposure and is shown in Figure \ref{fig:spec_J03384}. The spectra are immediately reminiscent of a WD, with a rising blue continuum, and broad H$\beta$ and H$\gamma$ absorption lines, however, no absorption is seen for H$\alpha$. The next striking feature in the spectra is the very strong and broad H$\beta$ emission core, while H$\alpha$ has a comparatively weak and narrow emission core, and is highly variable between the three sub-exposures. The WD nature of the system is further supported by its location in the colour-magnitude plot, Figure \ref{fig:CM_J03384}. All three sub-spectra also reveal possible weak cyclotron emission between $\lambda\lambda$ $\sim$ 5700\AA -- 6200\AA. Further evidence of the magnetic nature of the system is reported by Hernández-Díaz et al. (submitted), who determined a magnetic field of B = 9.6$^{+0.1}_{-0.5}$ MG from their analysis of the Balmer line Zeeman splitting. The system has an absolute Gaia G-band magnitude of G$_\text{abs}$ = 13.0 mag, and an X-ray luminosity of log(L$_\text{X}$) = 29.0 erg s$^{-1}$, further supporting the period bouncer interpretation for the system \citep[][]{period_bounce}

\section{Uncertain cases} \label{appen:uncertain_cases}

\subsubsection{Gaia DR3 3074742107676979584} \label{uncertain_1}

\begin{figure}[h!] 
  \centering
\begin{subfigure}[]{0.5\textwidth}
    \centering
\includegraphics[width=\linewidth]{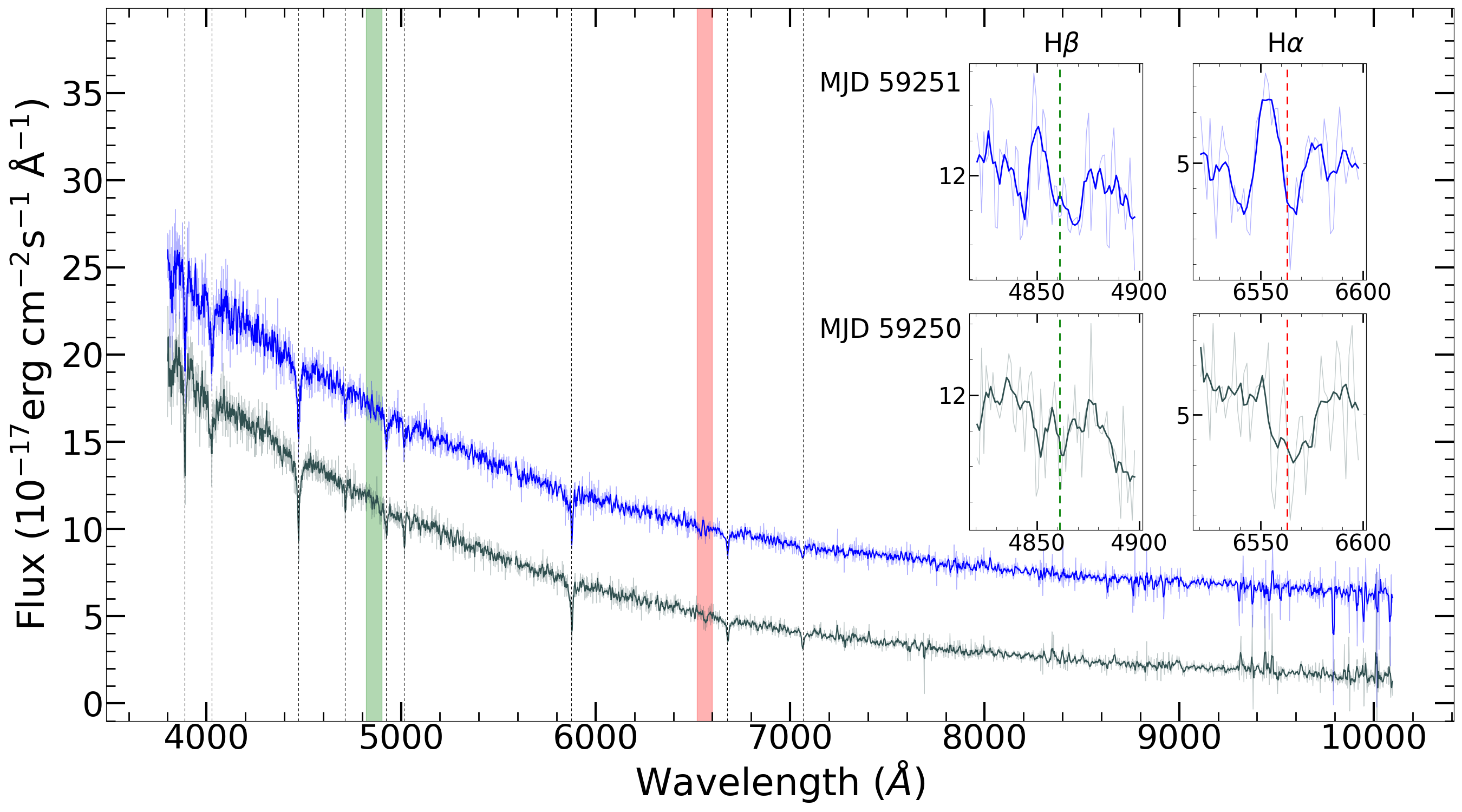}
         \caption{SDSS-V spectra} \label{appen:J085618_spectra}
\end{subfigure}%

\vspace{1ex}

\begin{subfigure}[]{0.5\textwidth}
    \centering
\includegraphics[width=\linewidth]{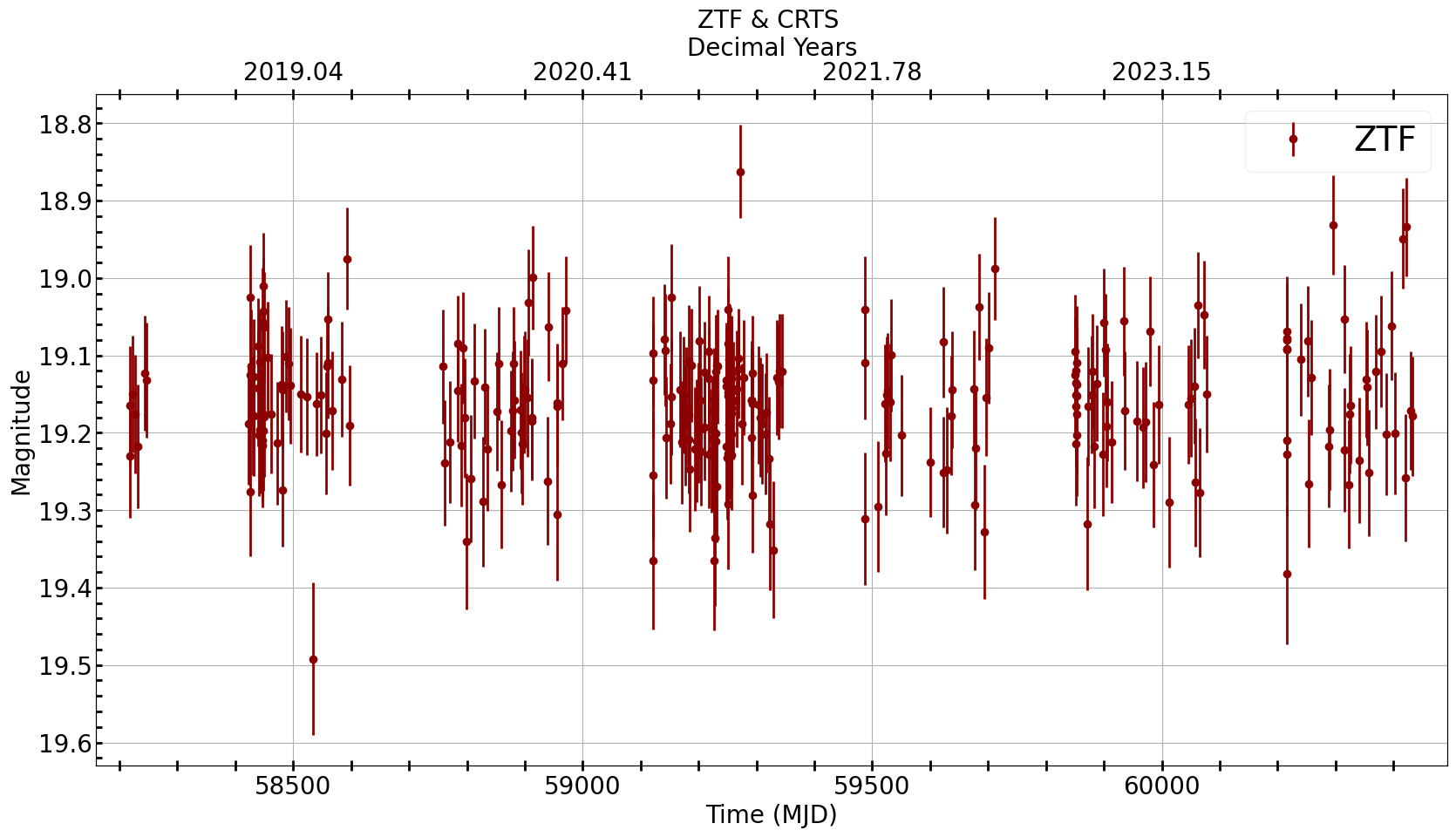}
  \caption{ZTF light curve}\label{appen:J085618_LC}
   \end{subfigure}  
 \begin{subfigure}[]{0.33\textwidth}
\includegraphics[width=\linewidth]{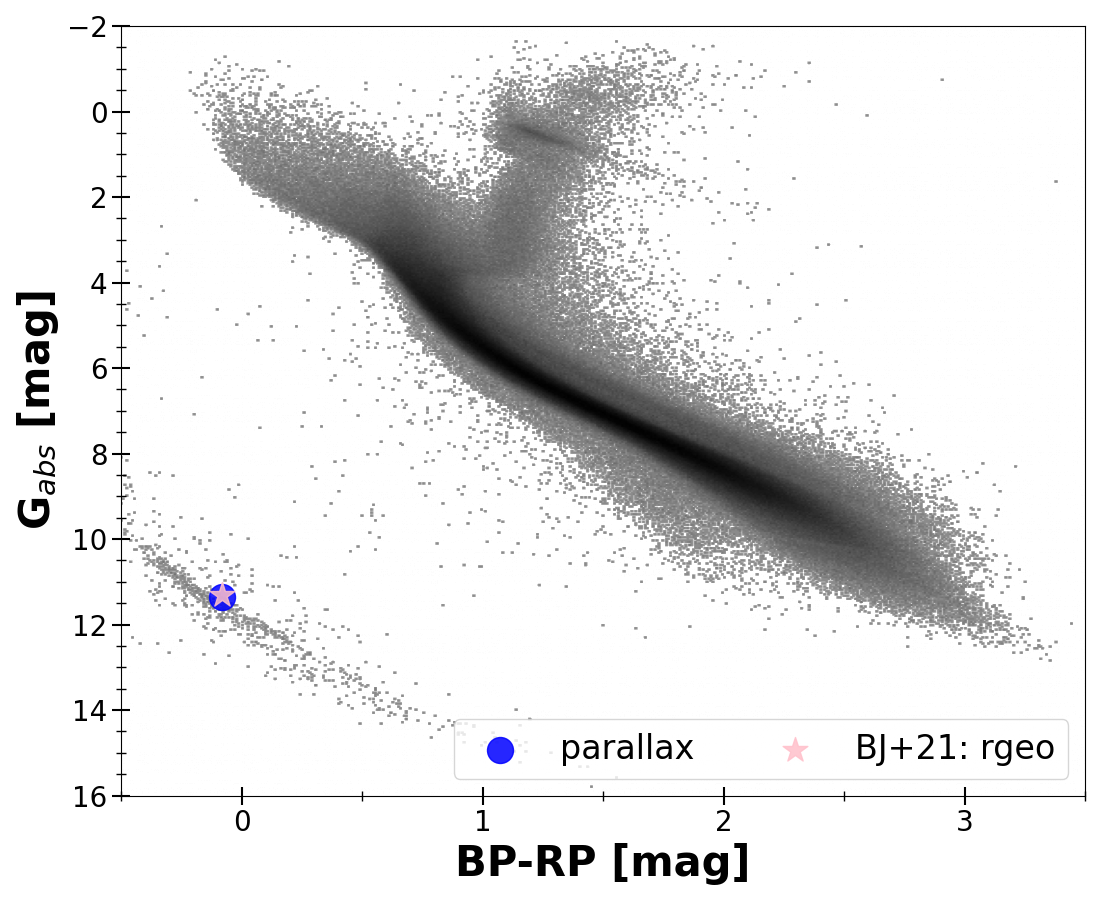}
         \caption{Colour-magnitude diagram}\label{appen:J085618_CM}
  \end{subfigure}

\caption{(a) Stacked spectra of Gaia DR3 3074742107676979584. The dark gray spectrum was obtained on MJD 59250, while the blue spectrum was obtained on MJD 59251. The MJD 59251 was shifted along the flux axis by 5 units to improve readability. Dashed black vertical lines indicate \ion{He}{i} absorption, while the red and green vertical strips highlights the areas around H$\alpha$ and H$\beta$, respectively, with the inset plots focusing on these regions. The left column of the inset plots shows the highlighted H$\beta$ region, with the green vertical dashed line indicating the H$\beta$ rest wavelength, while the right columns shows H$\alpha$, with the red vertical dashed line indicating the H$\alpha$ rest wavelength. The top row shows the spectrum obtained on MJD 59251, while the lower row shows the spectrum obtained on MJD 59250. The axes of the inset plots are the same as in the main plot. (b) Colour-magnitude diagram of Gaia DR3 3074742107676979584, indicating that the system lies on the WD track. (c) ZTF light curve of Gaia DR3 3074742107676979584, showing low amplitude variability.}
        \label{appen:J085618_plots}
\end{figure}

The first case we highlight is Gaia DR3 3074742107676979584. Looking at Figure \ref{appen:J085618_spectra}, which shows spectra from two consecutive nights, reveals the system to be a He WD (also known as DB WDs), with very weak H$\alpha$ absorption as well. This, in itself, is not that unusual, and in fact \citet[][]{He_WDs} found that H$\alpha$ absorption is seen in the majority, if not all, DB WDs, and are subclassified as DBA WDs. However, what is peculiar is the presence of weak, blue shifted, H$\alpha$ emission in the spectrum that was obtained on MJD 59251, which is not present the night before. It also appears that there might be very weak blue shifted H$\beta$ emission as well in both observations, being more prominent in the MJD 59251 observation, however this is might very well be noise in the spectrum. 

The presence of the H$\alpha$ emission line could suggest that very weak accretion is occurring in this system, and the transient nature of the emission line could suggest that the accretion might be phase dependent. The S/N-ratio for these observations were 14.77 and 12.07 for the MJD 59250 and MJD 59251 spectra, respectively, and thus we regard the H$\alpha$ emission feature to be real and not due to noise. However, a more likely explanation of the appearance of the H$\alpha$ emission could, however, simply be due to poor sky subtraction of the second observation. 

While Figure \ref{appen:J085618_CM} shows that the system indeed lies on the WD track, the ZTF light curve seen in Figure \ref{appen:J085618_LC} reveals a seeming photometrically stable system with very low-amplitude variability, however a few outlier events are seen in the light curve. There is no evidence in the spectrum of Gaia DR3 3074742107676979584 of magnetism, such as cyclotron harmonics or Zeeman-splitting, and thus a magnetic nature for the optical variability can be ruled out with the data at hand. Pulsations also seem unlikely as it would be difficult to explain the seemingly stable light curve. Another possibility might be a small second body in the system. This should be expected, as the vast majority of He WDs are thought to form due to binary interactions \citep[][]{He_WDs_population}. Now, the lack of any spectral features from a companion, and the position of the system in the colour-magnitude diagram (Figure \ref{appen:J085618_CM}) suggests that it has to be a very low-mass object and hence not a dwarf M-star. However, another, more likely, explanation is that the outliers are due to noise for a relatively faint object.
 
Thus, the transient H$\alpha$ emission seen in one of the spectra of Gaia DR3 3074742107676979584 might suggest that this is a weakly accreting He WD, however a more likely explanation might simply be due to poor sky subtraction, while the outliers in the ZTF light curve might be due to noise. It would, however, be worth follow-up spectroscopic observations to determine whether these conclusions are supported. 

\subsubsection{Gaia DR3 3134472183503345408} \label{uncertain_2}

\begin{figure}[h!] 
  \centering
\begin{subfigure}[]{0.5\textwidth}
    \centering
\includegraphics[width=0.9\linewidth]{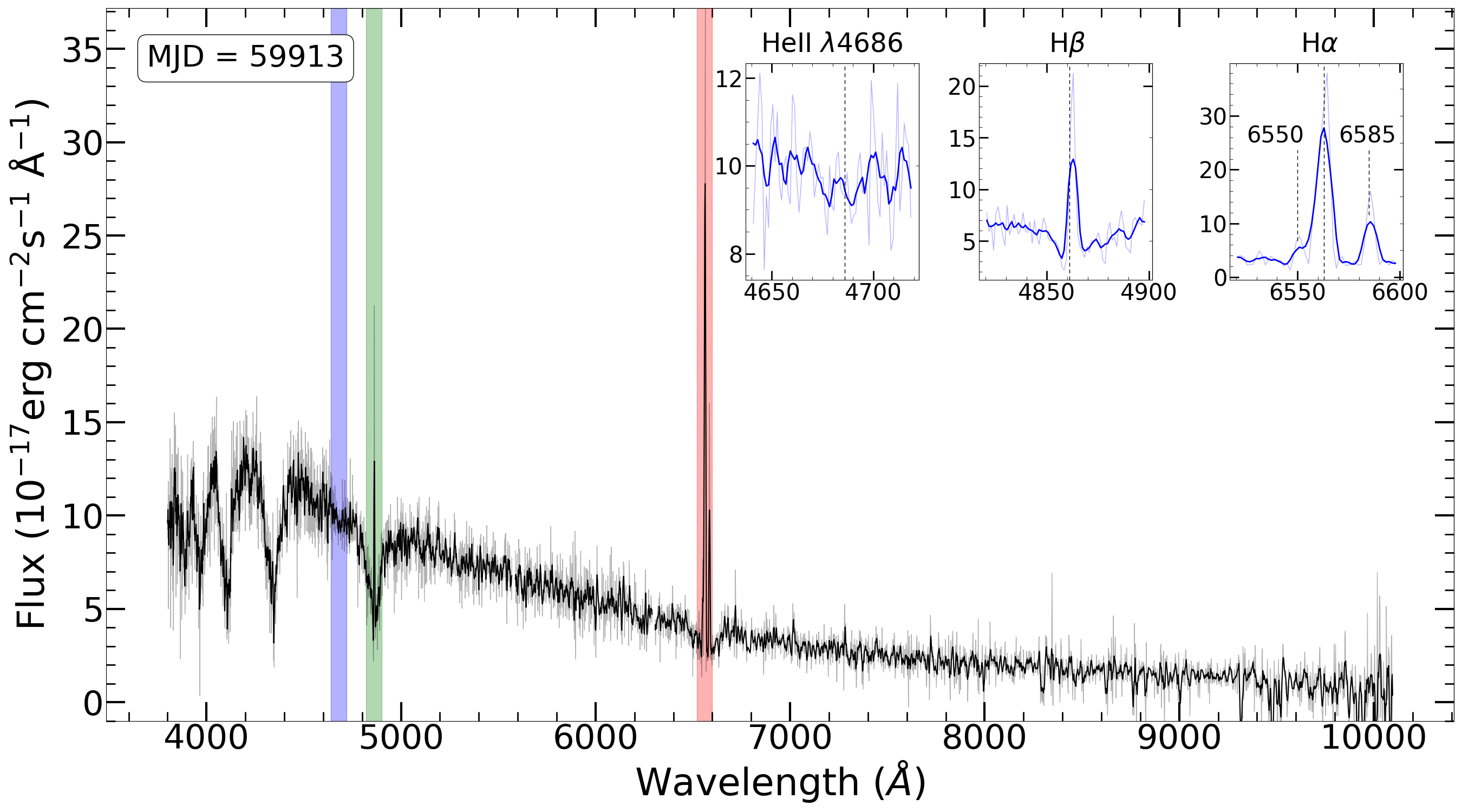}
         \caption{SDSS-V spectra}\label{appen:J064453_spectra}
\end{subfigure}%

\vspace{1ex}

\begin{subfigure}[]{0.5\textwidth}
    \centering
\includegraphics[width=0.9\linewidth]{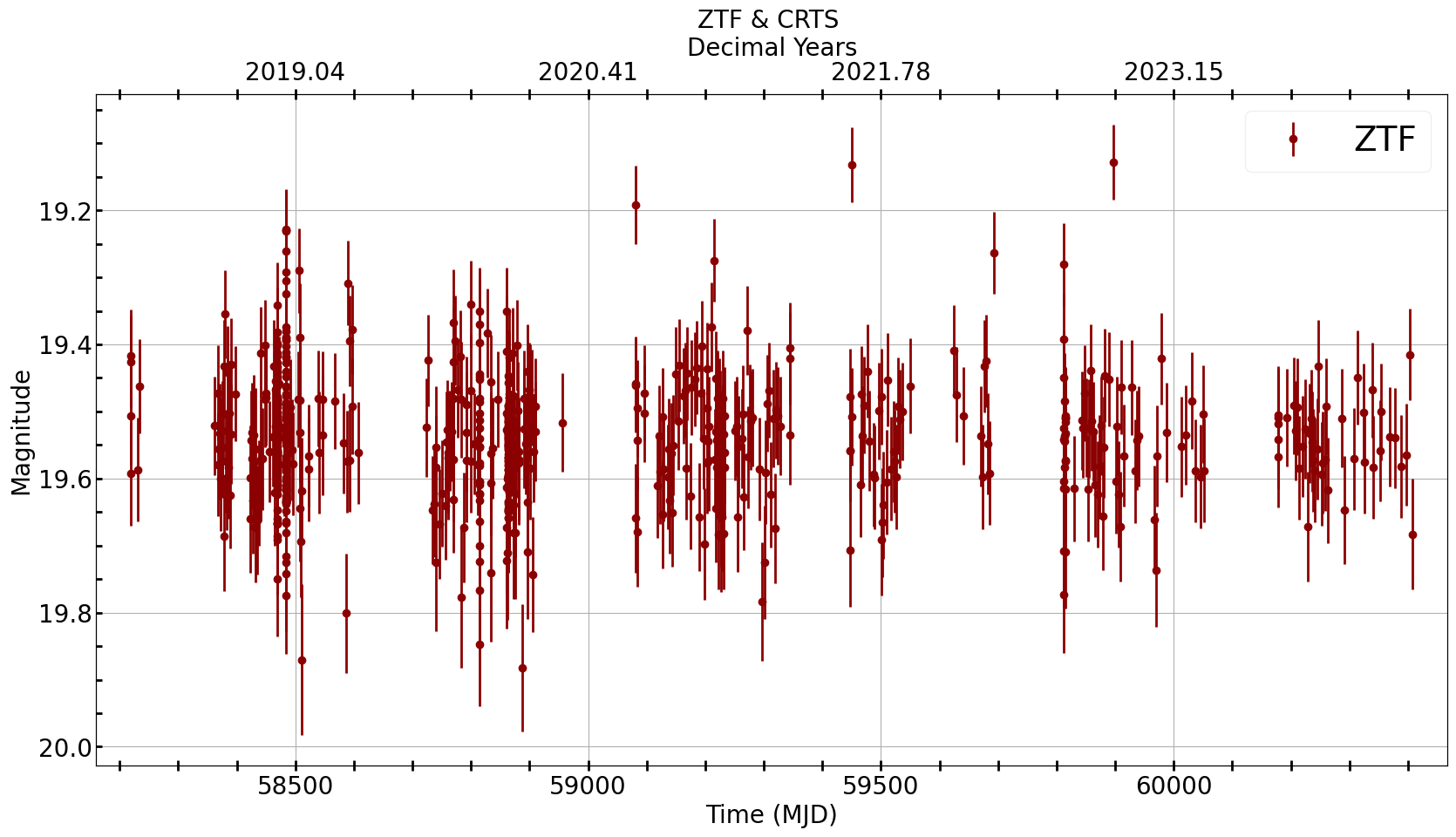}
  \caption{ATLAS light curve}\label{appen:J064453_LC}
   \end{subfigure}  
 \begin{subfigure}[]{0.33\textwidth}
\includegraphics[width=0.9\linewidth]{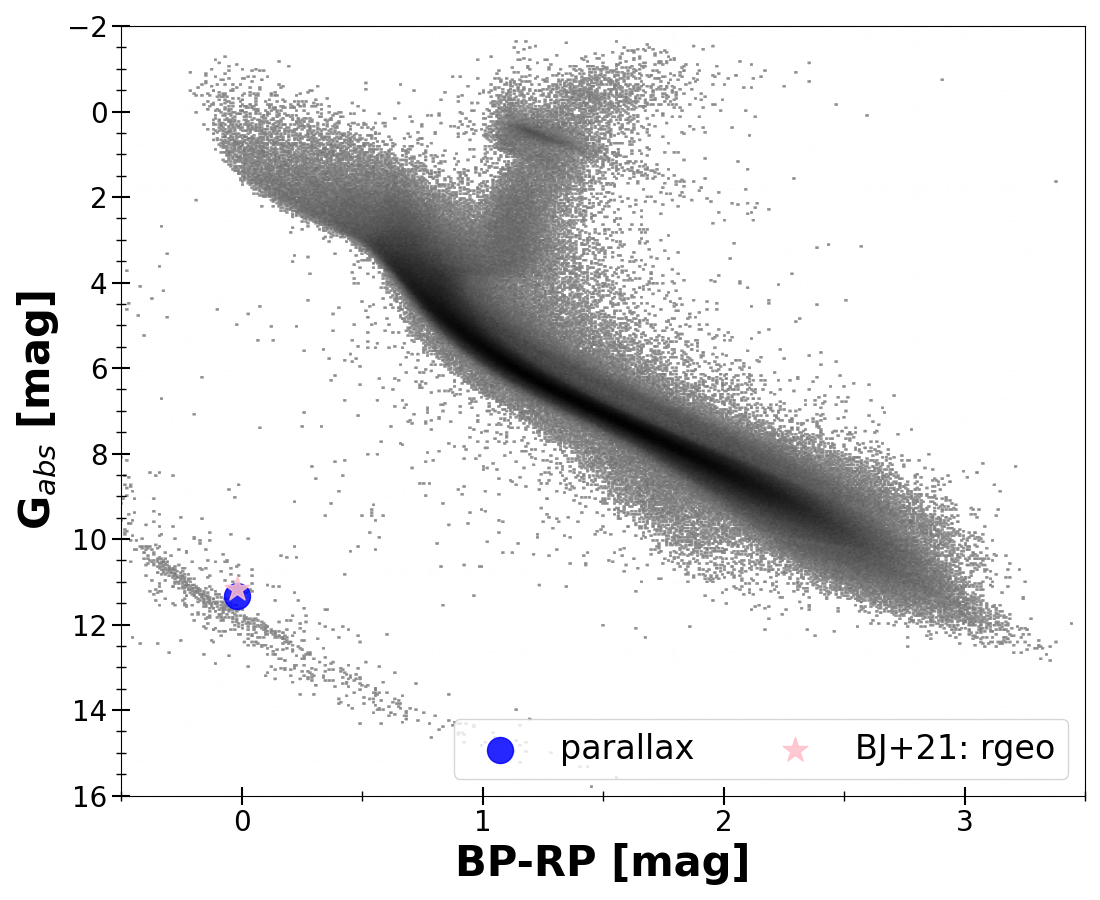}
         \caption{Colour-magnitude diagram}\label{appen:J064453_CM}
  \end{subfigure}

\caption{(a) Spectrum of Gaia DR3 3134472183503345408, with the position of likely [\ion{N}{ii}] emission is indicated in the H$\alpha$ inset plot. (b) Colour-magnitude diagram of Gaia DR3 3134472183503345408, indicating that the system lies on the WD track. (c) ZTF light curve of Gaia DR3 3134472183503345408.}
        \label{appen:J064453_plots}
\end{figure}

Gaia DR3 3134472183503345408 is another WD with peculiar spectral features. In Figure \ref{appen:J064453_spectra} we clearly see the broad Balmer absorption lines, indicative of a WD, while Figure \ref{appen:J064453_CM} further confirms that the system is a WD as it lies on the WD track. However, far from the clean Balmer absorption lines typically seen in isolated WDs, we see H$\alpha$ and H$\beta$ emission cores in the respective absorption lines. 

The spectrum shows no signs of magnetism, such as cyclotron humps or Zeeman-split absorption lines, nor does it show any signs of the presence of a companion, and the position on the colour-magnitude diagram would suggest that it is an isolated WD, or at least does not have a typical CV companion star. Thus, we can conclude that the emission is not from a coronally active companion, nor is it from the irradiated face of a typical main sequence star. Having a closer look at the H$\alpha$ region of the spectrum, the inset plot in Figure \ref{appen:J064453_spectra}, we notice that there is additional emission adjacent to H$\alpha$ at 6550\AA\ and 6585\AA. 

Even though Balmer emission line Zeeman splitting is very rare in isolated WDs \citep[][]{WD_balmer_emission_zeeman}, this is not what is happening in this system as the emission lines are not symmetric about H$\alpha$, and hence does not agree with what would be expected for linear Zeeman splitting. We are therefore left with concluding that the emission is from line emission, and attribute it to [\ion{N}{ii}] (with $\lambda_\text{rest}$ = 6548.4\AA\ and $\lambda_\text{rest}$ = 6583.4\AA, respectively). This introduces an additional question, what is the source of this emission, as these are not lines typically seen in WD spectra. One possibility is that the [\ion{N}{ii}] is from an old nova shell surrounding the WD \citep[][]{NII_from_novae}, however, there is no historical record of a nova associated with this system, nor is there anything in the spectrum that would suggest that the system might undergo nova eruptions, such as the presence of an accretion disk. 

However, visually inspecting the environment, the DSS2 colour and DSS2 red images reveals that the system lies in a nebulous region. We therefore inspected the SDSS spectra of other nearby objects and saw that some of those spectra (including those of objects SDSS\_ID 75089252 and SDSS\_ID 75089570) also showed the same [\ion{N}{ii}] emission lines. We also found another WD in the vicinity (SDSS\_ID 75089713) whose spectrum looked very similar to that of Gaia DR3 3134472183503345408, with Balmer emission cores. We therefore conclude that these emission lines are most likely due to the environment, and are present in the spectrum due to possible inadequate sky subtraction, however this needs to be investigated further. 

Photometrically the system appears to be variable, with the ZTF light curve, Figure \ref{appen:J064453_LC}, revealing the amplitude of variability to be $\sim$0.4 mag; however, a handful of brighter states are seen where the system brightens to $>$19.2 mag. The system also appears to have low states, where it falls to $<$ 19.8 mag. The nature of this variability remains unknown and also warrants further investigation. 

\section{Off target observations}

    \begin{figure}[h!]
     \centering
     \begin{subfigure}[b]{0.5\textwidth}
         \centering
         \includegraphics[width=\textwidth]{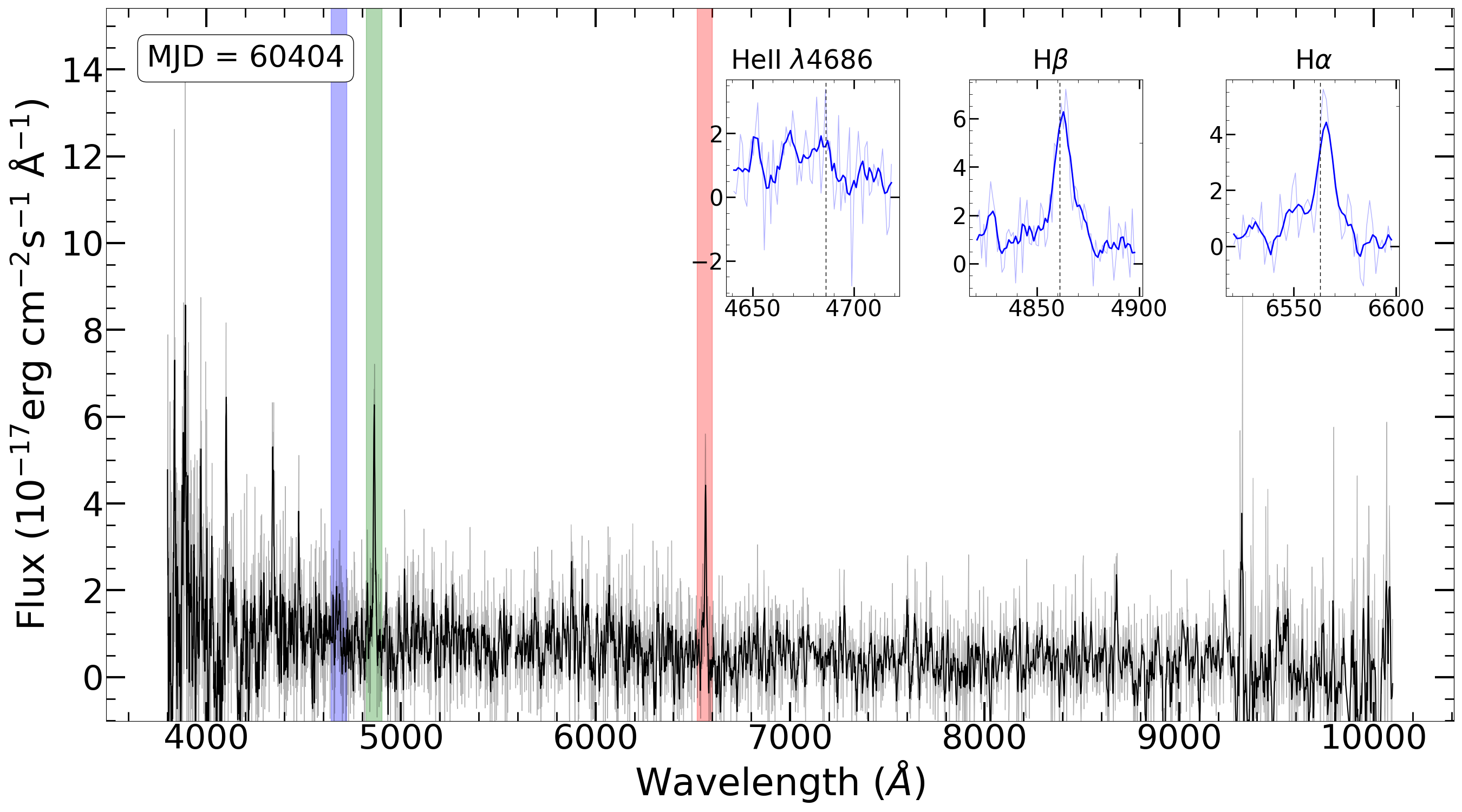}
         \caption{MJD 60404}
         \label{fig:appen_J135436_correct_positioning}
     \end{subfigure}

    \vspace{1ex}

     \begin{subfigure}[b]{0.5\textwidth}
         \centering
         \includegraphics[width=\textwidth]{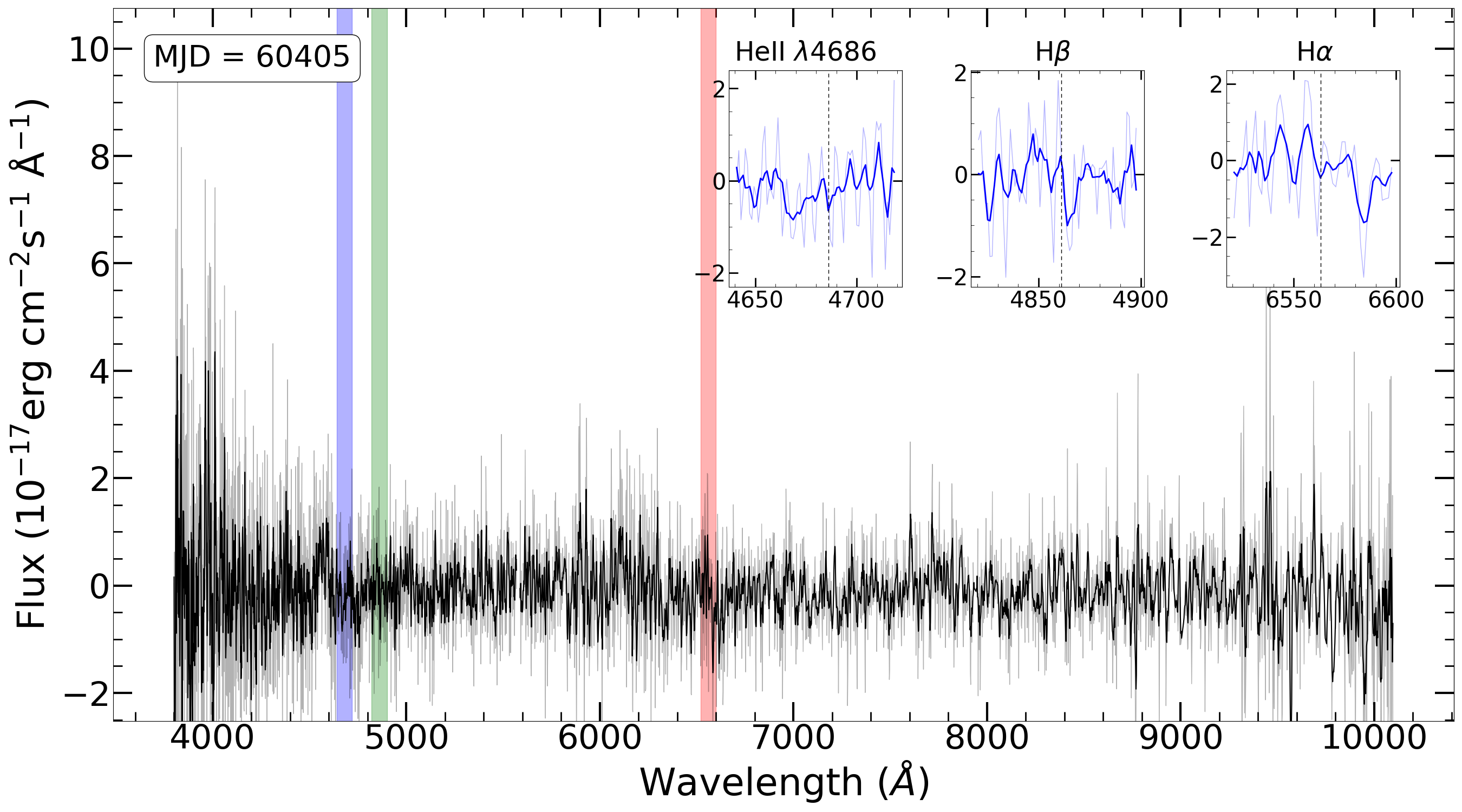}
         \caption{MJD 60405}
         \label{fig:appen_J135436_wrong_positioning}
     \end{subfigure}

    \vspace{1ex}

     \begin{subfigure}[b]{0.5\textwidth}
         \centering
         \includegraphics[width=0.8\textwidth]{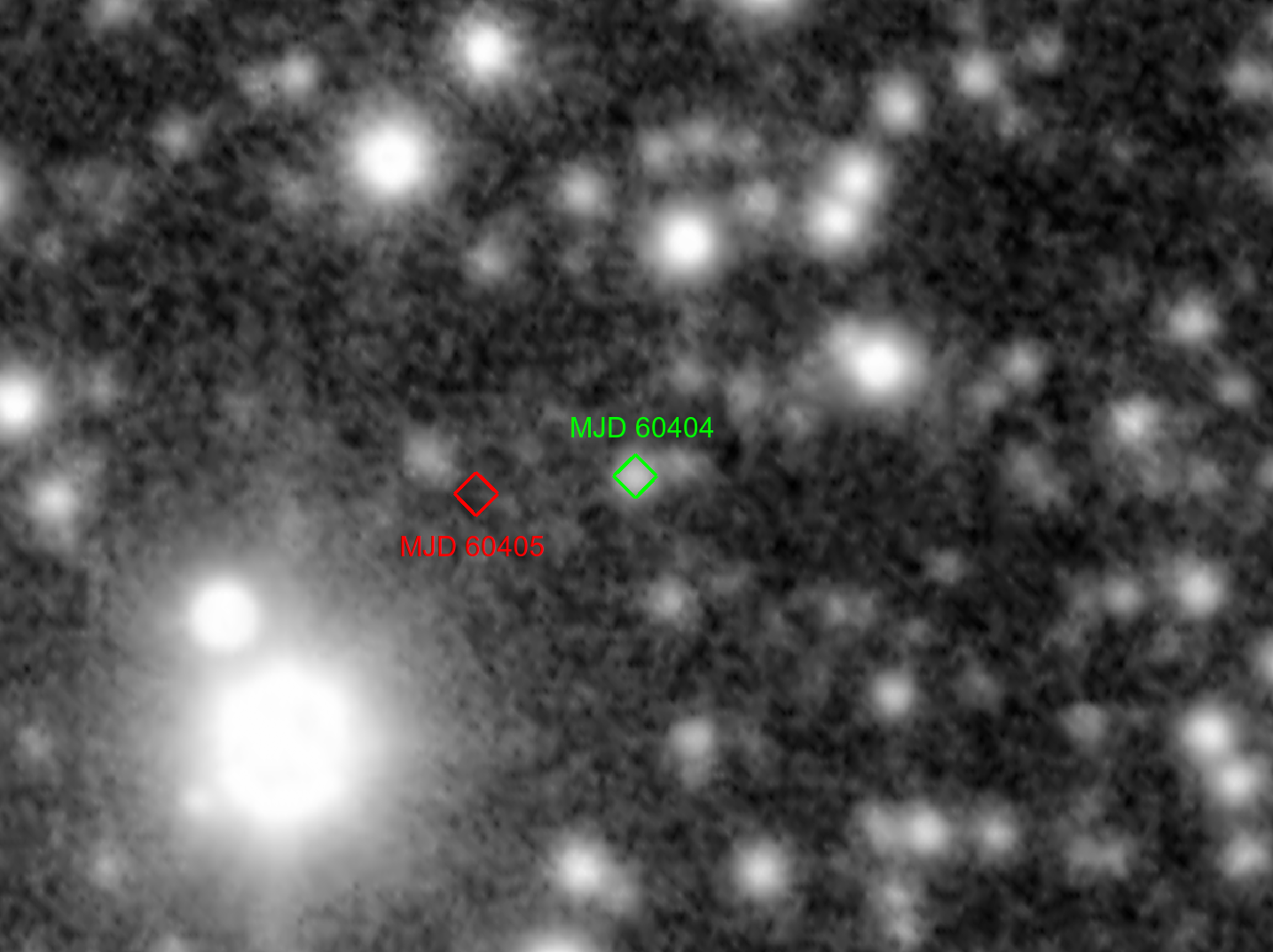}
         \caption{MJD 60404}
         \label{fig:appen_J135436_positioning_chart}
     \end{subfigure}
    
        \caption{Two spectra taken of 3eRASS J135436.2-610235 on different epochs, with the fibre positioning being on target on MJD 60404 (a), while being off target on MJD 60405 (b). In panels (c) we have DECaPS\_DR2 images, with the green and red diamonds indicating the fibre positions on MJD 60404 and MJD 60405, respectively.}
        \label{fig:appen_J135436_positioning}
    \end{figure}

\end{appendix}

\end{document}